\documentclass[12pt, draftclsnofoot, onecolumn]{IEEEtran}
\usepackage{bbding}
\usepackage{multirow}

\usepackage{setspace}
\usepackage{cite,graphicx,amsmath,amssymb}
\usepackage{fancyhdr}
\usepackage{mdwmath}
\usepackage{mdwtab}
\usepackage{balance}
\usepackage{xcolor}
\usepackage{bm}
\usepackage{amsthm}
\usepackage{threeparttable}

\usepackage{flafter}
\usepackage{makecell}
\usepackage{diagbox}
\usepackage{caption}
\usepackage{stfloats}

\usepackage{booktabs}

\usepackage{algorithm}  
\usepackage{algpseudocode}  
\usepackage{amsmath}

\usepackage{subcaption}

\usepackage{ulem}

\usepackage[T1]{fontenc}

\providecommand{\url}[1]{#1}
\allowdisplaybreaks
\allowdisplaybreaks[4]

\begin{document}
\title{Accelerating Mobile Edge Generation (MEG) by Constrained Learning}

\author{
  Xiaoxia Xu, 
  Yuanwei Liu, \IEEEmembership{Fellow,~IEEE,}
  Xidong Mu, 
  Hong Xing, 
  \\Arumugam Nallanathan, \IEEEmembership{Fellow,~IEEE}
  \vspace{-1.5em}

\thanks{X. Xu, Y. Liu, and A. Nallanathan are with the School of Electronic Engineering and Computer Science, Queen Mary University of
London, London E1 4NS, U.K. (email: \{x.xiaoxia, yuanwei.liu, a.nallanathan\}@qmul.ac.uk).}
\thanks{X. Mu is with the Centre for Wireless Innovation (CWI), Queen's University Belfast, Belfast, BT3 9DT, U.K. (x.mu@qub.ac.uk)}
\thanks{H. Xing is with the IoT Thrust, The Hong Kong University of Science and Technology (Guangzhou), Guangzhou, 511453, China, and she is also affiliated with the Department of ECE, The Hong Kong University of Science and Technology, HK SAR (e-mail: hongxing@ust.hk).}
}

\maketitle

\vspace{-1.5em}
\begin{abstract}  
    A novel accelerated mobile edge generation (MEG) framework is proposed for generating high-resolution images on mobile devices. 
    Exploiting a large-scale latent diffusion model (LDM) distributed across the edge server (ES) and the user equipment (UE), 
    only low-dimension features need to be transmitted to achieve cost-efficient artificial intelligence generative content (AIGC). 
    To reduce computation and transmission overheads, a dynamic diffusion and feature compression scheme is conceived. 
    By jointly optimizing the denoising steps and feature merging ratio, the image quality maximization problem is formulated 
    subject to latency and energy consumption constraints. To address this problem and tailor LDM sub-models, 
    a low-complexity protocol is developed. 
    Specifically, a backbone LDM architecture is first learned by offline distillation, which supports various compression options. 
    Then, dynamic diffusion and feature merging are predicted in online environment specific to channel observations. 
    To solve the resultant constrained Markov Decision Process (MDP) and overcome constraint violations, 
    a constrained variational policy optimization (CVPO) based MEG algorithm, \textit{MEG-CVPO}, is further proposed. Numerical results demonstrate that: 1) The proposed framework effectively improves image distortions, while reducing over $40\%$ latency compared to conventional generation schemes. 
    2) MEG-CVPO enables efficient constraint guarantees and realizes a flexible trade-off between image quality and generation costs.  
\end{abstract}

\section{Introduction}
\IEEEPARstart{O}{ver} the recent years, artificial intelligence generative content (AIGC) has emerged as a transformative technology 
to autonomously create diverse vivid contents ranging from images, videos, to music, thus revolutionizing the digital economy and information society.  
Leveraging the power of large-scale Generative Artificial Intelligence (GAI) models, such as ChatGPT \cite{ChatGPT}, Stable Diffusion, and DALL-E \cite{DALLE}, 
the high-fidelity contents generated by AIGC services have presented superior qualities over human artefacts. 
The autonomous and manipulable content creations and modifications can significantly reduce the man-power costs and time overheads, 
demonstrating high economic values \cite{GAI}. 
Driven by the burgeoning downstream digital applications of AIGC and the vision of connected intelligence,  
enabling AIGC on mobile devices, e.g., smartphones, has attracted arising attentions \cite{Mobile_GAI}. 
Interconnected with wireless radios, the on-device AIGC services provided by intelligent mobile networks are promising to support ubiquitous service access, human-machine interactions, and improved user experiences \cite{AIGC_SematicCom}. 

Due to the restricted computing and storage capacities of mobile devices \cite{DeviceOnlyInference}, 
the large-scale GAI models with billions of parameters and extensive computations are typically deployed on high-performance cloud servers. 
However, remote cloud services result in high latency, which limits the application of AIGC services on mobile applications. 
To reduce the latency in accessing mobile AIGC services, mobile edge generation (MEG) \cite{MEG} has been recently 
proposed as a promising solution for pulling down centralized generative models toward edge AI. 
The key idea is to enable distributed and scalable AIGC capabilities across multiple computation nodes in mobile edge network, which provides the following appealing benefits: 
\begin{itemize}
    \item MEG decomposes the sophisticated large-scale GAI models into distributed sub-models that are separately 
    deployed at user equipment (UE) and the nearby edge server (ES). 
    By jointly executing the distributed sub-models, 
    the edge resources can be fully exploited. 
    \item In conventional centralized generation, the server-generated contents need to be transmitted to the UE, which may lead to leakage of sensitive and personal information, and cause excessive service latency for high-resolution image or video generation.
    In MEG, the ES and UE only needs to exchange low-dimension features extracted by distributed GAI sub-models, 
    which reduces transmission latency and preserves users' privacy. 
    \item The GAI sub-model deployed at the UE can be customized for personalized applications and individual purposes, thus facilitating the personalization of AIGC services. 
    \item In addition, 
    efficient training and inference of GAI models can be supported
    by the beyond fifth-generation (B5G) and the sixth-generation (6G) techniques with massive connectivity, low latency, and bit-beyond transmissions.
\end{itemize}
Given the above attractive benefits,  MEG will contribute to the development and user experience improvement for mobile AIGC services.
To pave the way toward cost-efficient, latency-sensitive, and energy-constrained on-device AIGC, 
this paper investigates effective solution to accelerate MEG processing and unleash its potentials.

\subsection{Related Works}

\subsubsection{Related Works on Edge AI}
MEG can be regarded as a customized edge AI solution for AIGC services. 
In previous work, extensive research efforts have been devoted to edge AI for mobile deep neural networks (DNNs) 
applications, 
mainly including \textit{edge inference} and \textit{device-edge co-inference} \cite{Edge_Coinference_Li}. 
Specifically, \textit{edge inference} deploys DNNs on the ES in close proximity to mobile devices \cite{EdgeAI_Shi,EdgeAI_Xing}, 
thus avoiding high latency for routing data to the cloud and achieving fast inference. 
Nevertheless, uploading large-volume data to the ES, such as 3D images and high-resolution videos, still results in high transmission latency, which is intolerable for real-time mobile DNN applications. 
To combat this drawback, the \textit{device-edge co-inference} has been considered as another promising option \cite{Edge_Coinference_Li,Edge_Coinference_Kang}, 
which harnesses distributed computing resources across both ES and mobile devices to reduce the transmission overheads. 
The \textit{device-edge co-inference} can be simply implemented by splitting a pre-trained DNN into separate sub-models and deploys them over the ES and the UE, respectively \cite{Edge_Coinference_Spliting}. 
To further achieve feature compression for co-inference,  
a learning-based communication scheme was proposed in \cite{FeatureEncoding} to jointly optimize feature extraction, source coding, and channel coding. 
Based on information bottleneck framework \cite{InformationBottleneck}, a variational feature compression method was developed to adaptively identify and to prune redundant neurons of the encoded features. 
By integrating joint source-channel coding (JSCC), the authors of \cite{Bottlenet} jointly optimized the feature pruning and encoding in an end-to-end framework for image classification.

\subsubsection{Related Works on Mobile AIGC and MEG}

To support AIGC services at mobile edge network, the authors of \cite{Mobile_GAI} conceived a
 collaborative cloud and edge infrastructure. 
By performing AIGC model pre-training at cloud server and offloading content generation to edge devices, 
low-latency and personalized AIGC services can be achieved. 
To overcome limited resources and unstable channels, 
a semantic communication empowered AIGC generation and transmission framework was proposed in \cite{AIGC_SematicCom} to adjust edge and local computations. 
In \cite{MEG}, the authors introduced the concept of MEG and proposed various distributed deployment schemes for GAI models, 
thus reducing users' queueing latency in accessing GAI services. 
Furthermore, the authors of \cite{MEG_DT} proposed a novel diagram of MEG enabled digital twins and conceived single-user and multi-user generation mechanisms. 
Considering a multi-user collaborative diffusion model in wireless networks, 
the authors  of \cite{DistributedDiff} proposed a user-centric interactive AI method.  
By sharing several denoising steps for users with semantically similar prompts, 
energy constraints can be ensured while maximizing users' quality of experience (QoE). 
To deal with limited bandwidth resources and dynamic channels at mobile edge network, 
a pricing-based incentive mechanism was developed in \cite{AIGC_Pricing} 
for AIGC generation and transmission, thus maximizing the utility of users. 
Furthermore, focusing on the security aspects, 
the authors of\cite{TrustWorthAIGC} proposed a novel paradigm named TrustGAIN to 
deliver trustworthy AIGC services in 6G networks, which can efficiently defend against malicious or fake messages.

\subsection{Motivations and Contributions}
Previous studies have laid a solid foundation for training and inference of sophisticated AI models at mobile edge network. 
However, effective MEG architectures and resource-constrained solutions for low-latency on-device generation are still less explored currently. 
Due to the fundamental trade-off between content generative quality and resource limitations,   
there are several crucial challenges to meet stringent AIGC service requests and to accelerate MEG: 
\begin{itemize}
\item The emerging GAI models (e.g., diffusion model (DM) and Transformer) 
rely on tailored compression schemes at mobile edge network to reduce the end-to-end generation latency, which requires different designs compared to 
conventional DNNs such as multi-layer perceptron (MLP) and convolutional neural networks (CNN). 
\item To reduce transmission overheads, conventional edge AI typically exploits neuron pruning to compress transmitted features for classification/regression. 
However, the generation qualities of AIGC are very sensitive to neuron pruning, particularly for high-resolution data such as image and video streams. 
To achieve low-latency generation, effective feature compression strategies should be investigated for MEG. 
\item Current GAI model compression techniques mainly adopt static compression methods for computational acceleration, which is limited to pre-determined configurations and commonly ignores transmission overheads. 
However, the on-device generation should satisfy end-to-end latency and energy consumption constraints adaptive to channel conditions in an online fashion. 
To enable controllable distortion-latency trade-off, efficient constrained learning methods are required for dynamic end-to-end acceleration of MEG.
\end{itemize}

To address the aforementioned issues, we propose a novel accelerated MEG framework for high-resolution image generation in this paper. 
The proposed framework distributes the large-scale GAI model, i.e., latent diffusion model (LDM), across both the ES and the UE. 
To accelerate joint execution of distributed sub-models, a dynamic diffusion and feature merging scheme is proposed, 
which achieves few-step diffusion and length-adaptive transmitted features to relieve computation and communication overheads, respectively. 
Different from conventional pruning-based feature transmission schemes that discard information of pruned neurons, 
the proposed scheme merges similar neurons to construct low-dimension features, 
thereby retaining content details. 
Our goal is to maximize the image generation quality while satisfying latency and energy consumption constraints. 
This is formulated as a high-complexity non-convex optimization problem, where the objective function is a black-box function without explicit mathematical modelling.  
To make it tractable, a low-complexity dynamic MEG acceleration protocol is devised. 
A backbone architecture for LDM sub-models is first learned through the offline distillation. 
Then, the denoising steps and feature merging ratios are dynamically determined in the online prediction by solving a constrained Markov Decision process (MDP). 
To achieve constrained learning for MEG acceleration, we further develop a constrained variational policy optimization (CVPO) based MEG algorithm, namely MEG-CVPO. 
The proposed MEG-CVPO algorithm can improve the dynamic compression policy over a trusted region that results in feasible policy distributions. 
Our main contributions can be summarized as follows. 
\begin{itemize}
    \item We propose a novel accelerated MEG framework, which 
    enables cost-efficient on-device generation of high-resolution images via a large-scale LDM distributed across the UE and the nearby ES.  
    By jointly optimizing the denoising steps and feature merging ratios, the image generation quality is maximized subject to both latency and energy constraints.  
    \item We develop a low-complexity protocol for dynamic MEG acceleration. 
    The designed protocol first learns a backbone architecture via the offline distillation. 
    Based on the backbone architecture, on-demand dynamic diffusion and feature merging can be realized in the online channel environment. 
    This recasts the formulated problem as a constrained MDP.
    \item We propose a constrained reinforcement learning (RL) algorithm for MEG by invoking the CVPO theory,  namely MEG-CVPO. 
    Relying on the variational inference, the policy is trained to improve system rewards while achieving feasible policy distributions. 
    This effectively mitigates constraint violations and solves the constrained MDP. 
    \item We provide numerical results to verify the effectiveness. 
    Compared to conventional generative schemes, the proposed dynamic diffusion and feature merging scheme can reconstruct high-quality $1024\times 1024$ 
    images over noisy channels within $3\sim 7$ seconds, which achieves over $40\%$ latency reduction. 
    Furthermore, the MEG-CVPO algorithm effectively provides constraint guarantees compared to conventional Lagrangian-based learning methods, thereby striking a controllable trade-off between generation quality and costs. 
\end{itemize}

\subsection{Organization and Notation}
The rest of this paper is organized as follows. 
Section II presents the accelerated MEG framework and formulates the image generation quality maximization problem. 
In Section III, a low-complexity protocol for dynamic compression is designed. 
Furthermore, a constrained MEG-CVPO learning algorithm is developed in Section IV to achieve dynamic diffusion and feature merging solutions. 
Section V provides numerical results to verify efficiency of the proposed MEG framework. 
Finally, Section VI concludes the paper.

\textit{Notations}: Vectors and matrices are denoted by bold-face letters; 
$\lfloor x \rceil$ indicates the rounding operation of variable $x$;  
$\left[x\right]^{+} = \max\left\{x,0\right\}$; 
$x \propto y$ indicates that variable $x$ is proportional to variable $y$; 
and $\|\mathbf{x}\|$ denotes the Euclidean norm of a vector $\mathbf{x}$.

\section{System Model and Problem Formulation}

\subsection{An Accelerated MEG Framework}
\begin{figure*}[!htbp]
    \centering
    \includegraphics[width=1\textwidth]{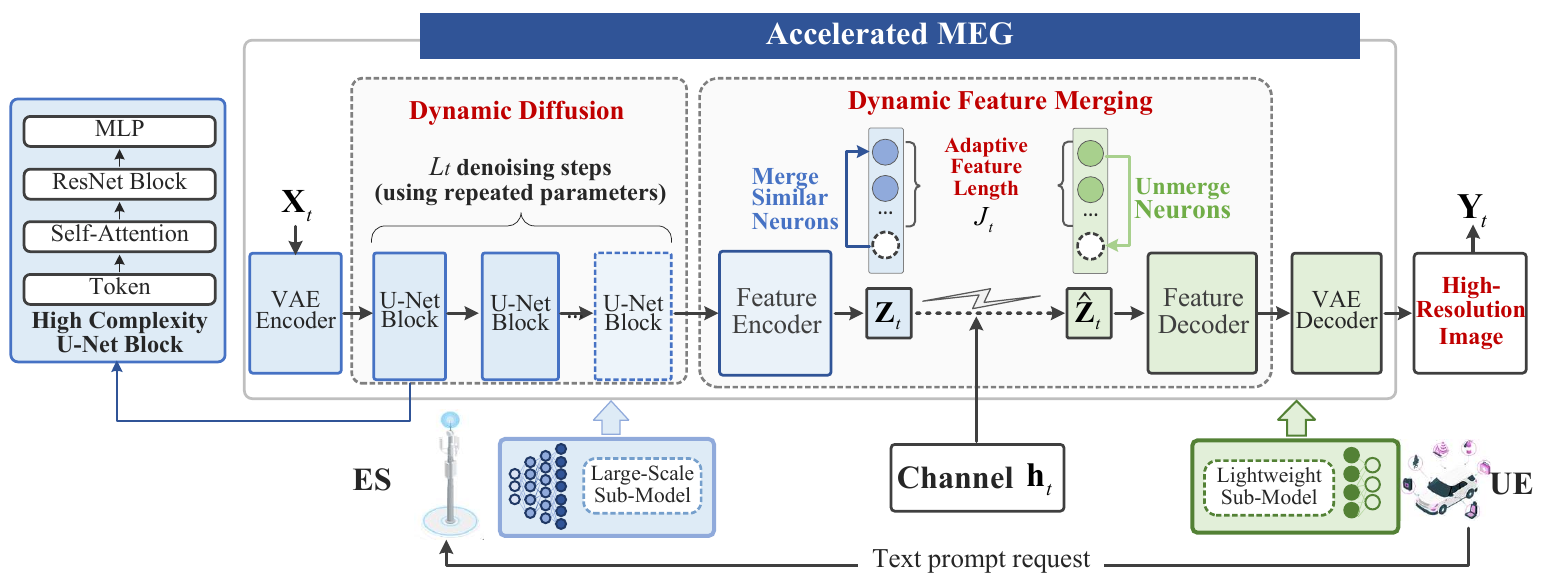}
    \caption{The proposed dynamic MEG acceleration framework.}
    \label{fig_MEG_framework}
\end{figure*}

To enable real-time AI content generation on mobile devices, 
we propose an accelerated MEG framework empowered by edge intelligence environment, as shown in Fig. \ref{fig_MEG_framework}. 
Without loss of generality, the MEG framework adopts the typical point-to-point transmission for mobile edge communication,  
where an ES equipped with a high-performance computing platform can communicate with a low-cost UE to offer  AIGC services for mobile applications. 
The ES sequentially handles a total number of $T$ image generation tasks 
offloaded by the UE in $T$ time frames, each corresponding to one task, indexed by $\mathcal{T}=\left\{1,2,...,T\right\}$. 

\subsubsection{MEG Basics}
Due to limited computational power and memory of the mobile terminal, it is generally difficult for the UE 
to handle the intensive computation workloads required by GAI model. 
To address this issue, MEG decomposes the original GAI model into two associated sub-models, 
which encompass a large-scale sub-model deployed at the ES and a tiny sub-model at the UE, respectively.
By doing so, AIGC can be achieved by collaborative execution of  the decomposed sub-models. 
Upon receiving the generation service requests from the UE, 
the ES first performs computations involving high-complexity DNN modules, e.g., self-attention Transformers, via powerful graphics processing units (GPUs).  
Then, important features from the intermediate results can be extracted and encoded into low-dimension features, 
which are sent to the UE through noisy wireless channels. 
By decoding the received features, the UE further performs high-quality content generation locally via lightweight computation.

\subsubsection{Compression Schemes of MEG}
To accelerate on-device generation in the mobile edge environment, a crucial aspect is the cost-effective compression of both large-scale GAI sub-models and transmitted features. 
In this work, we propose a novel dynamic compression structure for LDM. 
LDM is one of the most popular GAI models in current AIGC applications for text-to-image/image-to-image generations. 
We focus on text-to-image generation here. 
As shown in Fig. \ref{fig_MEG_framework}, a standard LDM typically consists of three components, i.e., a VAE encoder, a denoising diffuser based on U-Net \cite{UNet}, and a VAE decoder.  
Since each U-Net block includes multiple high-complexity self-attention Transformer layers and ResNet layers, 
the denoising diffusion process usually requires intensive computations. 
Considering limited computing resources of UE, 
we deploy the VAE encoder and the high-complexity U-Net denoising diffuser at the ES, while implementing the lightweight VAE decoder at the UE, respectively. 
After receiving the text prompts from users, the ES encodes and denoises the latent features via the VAE encoder and denoising diffuser. 
By transmitting the extracted latent features to the UE, high-resolution images are then decoded by the VAE decoder. 
Therefore, the intensive computations of LDM can be achieved by collaborative inference of the ES and the UE.

Since the ES and the UE only exchange low-dimension latent features instead of high-dimension raw images, 
MEG reduces the service latency compared to conventional centralized generation schemes. 
Nevertheless, the denoising diffuser repeatedly executes a U-Net block for tens of denoising steps to maintain performance \cite{denoising,DDIM}, 
which usually requires a time-consuming complex reverse diffusion process. 
Furthermore, the latent features extracted by the ES-side sub-model need to be transmitted through noisy channels to the UE-side for image decoding, 
which still incurs communication overheads and quality degradation. 
To further alleviate computational and transmission burdens, our proposed compression structure advocates 
a dynamic reverse diffusion process for generation acceleration and a variable-length feature merging 
for efficient transmissions.
\begin{itemize}
    \item \textbf{Dynamic Diffusion}: We define $L_{\max}$ as the maximal number of denoising steps at the denoising diffuser, 
    which reuses the U-Net block parameters to execute 
    self-attention and convolutional operations, thus denoising and extracting the latent features for content generation. 
    We define an accelerating ratio  $\alpha_{t}\in[0,1]$ to decide the time duration of the diffusion process at each time frame $t$, 
    where only $L_t=\lfloor \alpha_{t} L_{\max} \rceil$ denoising steps will be performed. 
    \item \textbf{Feature Merging}:
    Given generation task $g_{t}$, 
    the ES exploits the VAE encoder and dynamic reverse diffusion process to obtain the 
    latent feature $\mathbf{Z}_{t}^{\mathrm{E}} =\left[\mathbf{z}_{1,t}^{\mathrm{E}},\mathbf{z}_{2,t}^{\mathrm{E}},\dots,\mathbf{z}_{J_{\max},t}^{\mathrm{E}}\right] 
    \in \mathbb{R}^{d^{\mathrm{C}}\times J_{\max}}$. 
    Moreover, $d^{\mathrm{C}}$ and $J_{\max}$ denote the original channel number and length of the feature extracted by the ES-side sub-model, respectively. 
    The latent feature will be transmitted from the ES to the UE to perform on-device high-quality content generation. 
    To accommodate feature transmission overheads and latency in accordance with the restricted environment of the wireless system,  
    we develop a novel feature merging scheme to dynamically decrease the feature length. 
    Specifically, by determining the merging ratio $\beta_{t}$ at each time frame, 
    $\lfloor\beta_{t}J_{\max}\rceil$ vectors will be trimmed,  
    and their information will be merged into the remaining neurons. 
    This leads to a length-adaptive feature $\mathbf{Z}_{t}\in\mathbb{R}^{d^{\mathrm{C}}\times J_{t}}$ with length $J_{t} = J_{\max} - \lfloor\beta_{t}J_{\max}\rceil$.  
\end{itemize}

The detailed implementations of the dynamic diffusion and feature merging schemes will be discussed in Section \ref{Sec_DynamicDiff}.

\subsection{Computation and Transmission Models}
\begin{figure*}[!b]
    \vspace{1em}
    \centering
    \includegraphics[width=1\textwidth]{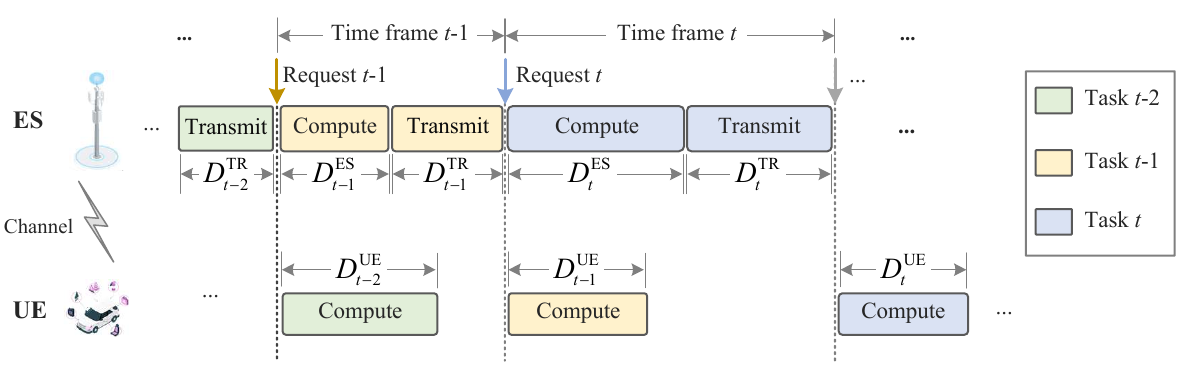}
    \caption{A pipeline of the proposed MEG for ES-UE co-inference.}
    \label{fig_slot}
\end{figure*}

At each time frame $t$, the ES will fetch a generation task request (i.e., the text prompt request) uploaded by the UE in the previous time frame. 
Since the text prompt upload typically consumes minor bandwidth and power resources, we ignore the text prompt uploading overheads here.
As shown in Fig. \ref{fig_slot}, given the input $\mathbf{X}_{t}$ of task $g_{t}$ in time frame $t$, 
the latency and energy consumption for completing a generation task depend on 
both the computation and transmission overheads, which will be modelled in the sequel, respectively. 

\subsubsection{Computation Model}

The computational latency at the ES is decided by floating point operations (FLOPs) of the VAE encoder, FLOPs of the U-Net block in the denoising diffuser, 
and the denoising steps, i.e., 
\begin{equation}
        D_{t}^{\mathrm{ES}}\left(\alpha_{t}\right) = \frac{O^{\mathrm{E}} + \lfloor \alpha_{t}L_{\max} \rceil O^{\mathrm{UNet}}}{f^{\mathrm{ES}}} + D_{0}^{\mathrm{ES}}, 
\end{equation}
where $O^{\mathrm{E}}$ and $O^{\mathrm{UNet}}$ indicate the required FLOPs for computing the VAE encoder and a single U-Net block, respectively.  
$D_{0}^{\mathrm{ES}}$ is the average latency for accessing the learned GAI sub-models. 
Moreover, $f^{\mathrm{ES}}$ denotes the floating point operation per second (FLOPS) of ES's GPU that measures the data processing speed. 
Similarly, the computational latency at UE can be given by
\begin{equation}
        D_{t}^{\mathrm{UE}}=O^{\mathrm{D}}/{f^{\mathrm{UE}}}+D_{0}^{\mathrm{UE}}, 
\end{equation}
where $O^{\mathrm{D}}$ denotes the FLOPs of VAE decoder, 
$D_{0}^{\mathrm{UE}}$ represents the average latency for accessing the learned decoder at the UE,  
and $f^{\mathrm{UE}}$ is the FLOPS at the UE's processor.

On the other hand, the computational energy consumption can be estimated by models' FLOPs and the energy efficiency of the computing hardware \cite{Energy_FLOPs}. 
Based on the dynamic diffusion model, the computational energy consumption for the proposed framework can be given by
\begin{equation}
    E_{t}^{\mathrm{ES}}\left(\alpha_{t}\right) = 
    \frac{O^{\mathrm{E}} + \lfloor \alpha_{t}L_{\max} \rceil O^{\mathrm{UNet}}}{\eta^{\mathrm{ES}}}
    + E_{0}^{\mathrm{ES}},     
\end{equation}
where $\eta^{\mathrm{ES}}$ denotes the energy efficiency of ES computing, 
i.e., the energy dissipations for each float-operation point computation, 
and $E_{0}^{\mathrm{ES}}$ is the average energy consumption for GAI sub-model access.

\subsubsection{Transmission Model}
The transmission latency depends on both the size of the transmitted feature and the channel gain of UE. 
    Hence, the transmission latency can be modelled by 
    \begin{equation}
        D_{t}^{\mathrm{TR}}\left(\beta_{t}\right)=
         \frac{b d^{\mathrm{C}} \left(J_{\max} - \lfloor \beta_t J_{\max} \rceil\right)
         + J_{t}^{\mathrm{aux}}}{B_{0}\log_{2}\left(1+\frac{P_{0}h_{t}}{B_{0}N_{0}}\right)},
    \end{equation}
where $h_{t}$ denotes the UE's channel gain at time frame $t$,  
$b$ is the number of bits of each floating number,  and $B_0$ indicates the bandwidth of each resource block (RB). 
$P_{0}$ denotes the downlink transmit power of the ES, and $N_{0}$ is the additive white Gaussian noise (AWGN).
Moreover, $J_{t}^{\mathrm{aux}}$ represents the overheads of additional information required to be transmitted to UE in order to keep the dimension consistent between the features and the UE-side sub-model. 
For feature pruning, UE can acquire the neuron importance in advance and pad the pruned neurons with zero values, which leads to $J_{t}^{\mathrm{aux}}=0$.
By comparison, as feature merging is task-specific, an additional information overhead $J_{t}^{\mathrm{aux}}=b\lfloor \beta_t J_{\max} \rceil$ is required 
to inform UE which dimensions the pruned neurons have been merged into, and thereby enable the feature unmerging at the UE.

Meanwhile, the energy consumption to transmit the encoded feature $\mathbf{Z}_{t}$ 
between the ES and the UE is expressed as
\begin{equation}
    E_{t}^{\mathrm{TR}}\left(\beta_{t}\right) = P_{0} D_{g,t}^{\mathrm{TR}}
    = \frac{P_{0} \left[b d^{\mathrm{C}} \left(J_{\max} - \lfloor \beta_t J_{\max} \rceil \right) + J_{t}^{\mathrm{aux}}\right]}
    {B_{0}\log_{2}\left(1+\frac{P_{0}h_{t}}{B_{0}N_{0}}\right)}. 
\end{equation}

To sum up, the end-to-end generation latency to accomplish task $g_{t}$ is given by
\begin{equation}
    D_{t}\left(\alpha_{t},\beta_{t}\right)
    =D_{t}^{\mathrm{ES}}\left(\alpha_{t}\right)+ D_{t}^{\mathrm{UE}} + D_{t}^{\mathrm{TR}}\left(\beta_{t}\right).
\end{equation}
On the other hand, the computing energy consumption for completing the generation task $g_t$ 
based on the proposed MEG framework is as follows:
\begin{equation}
    E_{t}\left(\alpha_{t},\beta_{t}\right)
    =E_{t}^{\mathrm{ES}}\left(\alpha_{t}\right) + E_{t}^{\mathrm{UE}}
    + E_{t}^{\mathrm{TR}}\left(\beta_t\right). 
\end{equation}

\subsection{Problem Formulation}
Based on the proposed framework, 
the generated contents $\mathbf{Y}_t$ 
can be expressed as a function of the diffusion accelerating ratio $\alpha_{t}$, feature merging ratio $\beta_{t}$, 
ES-side sub-model $\mathcal{F}^{\mathrm{E}}(\cdot)$ for feature encodeing, and UE-side sub-model $\mathcal{F}^{\mathrm{D}}(\cdot)$ for feature decoding, i.e.,
\begin{equation}
    \mathbf{Y}_t = \mathcal{Y}_t \left(\alpha_t,\beta_{t},\mathcal{F}^{\mathrm{E}}(\cdot),\mathcal{F}^{\mathrm{D}}(\cdot)\mid\mathbf{X}_{t},h_{t}\right). 
\end{equation} 
Therefore, we can formulate a dynamic compression optimization problem for accelerating MEG under limited latency and energy consumption. 
Our goal is to maximize the content generative qualities,  
denoted by $f_{\mathrm{G}}\left(\mathbf{Y}_{t}\mid\mathbf{X}_{t},h_{t}\right)$, $t\in\mathcal{T}$, 
by jointly optimizing the diffusion accelerating ratio $\alpha_{t}$, feature merging ratio $\beta_{t}$, and GAI sub-models $\mathcal{F}^{\mathrm{E}}(\cdot)$ and $\mathcal{F}^{\mathrm{D}}(\cdot)$, 
subject to the average end-to-end latency and energy consumption constraints.
The corresponding problem is formulated as
\begin{subequations}\label{P0}
\begin{align*}
\mathcal{P}_{0}:&\max_{\alpha_t,\beta_{t},\mathcal{F}^{\mathrm{E}}(\cdot),\mathcal{F}^{\mathrm{D}}(\cdot)} ~ \frac{1}{T}\sum\limits_{t\in\mathcal{T}}
f_{\mathrm{G}}\left(\mathbf{Y}_{t}\mid\mathbf{X}_{t},h_{t}\right) \tag{\ref{P0}{a}}
\\ {\mathrm{s.t.}}~ &
\mathrm{C}1:~\frac{1}{T}\sum_{t\in\mathcal{T}} D_{t}\left(\alpha_{t},\beta_{t}\right) \leqslant D_{\max},  ~ \forall t\in\mathcal{T}, \label{constraint_Latency} \tag{\ref{P0}{b}}
\\&
\mathrm{C}2:~\frac{1}{T}\sum_{t\in\mathcal{T}} E_{t}\left(\alpha_{t},\beta_{t}\right)
\leqslant E_{\max},  ~ \forall t\in\mathcal{T}, \label{constraint_Energy_UE} \tag{\ref{P0}{c}}\\&
\mathrm{C}3:~0 \leqslant \alpha_{t}, \beta_{t} \leqslant 1, ~ \forall t\in\mathcal{T}. \label{constraint_binary} \tag{\ref{P0}{d}}
\end{align*}
\end{subequations}

While problem $\mathcal{P}_0$ strikes a trade-off between generation quality and latency/energy consumption, 
there are several difficulties to search for the optimal solution. 
First, the highly non-convex length-adaptive neuron merging and dynamic diffusion is an NP-hard problem. 
Moreover, the objective function  $f_{\mathrm{G}}\left(\cdot\right)$ lacks explicit expression due to black-box operation of GAI models, 
and the end-to-end generation latency $D_t(\alpha_{t},\beta_{t})$ has unknown dynamics decided by time-varying channel realizations.
Furthermore, it is non-trivial to achieve on-demand MEG acceleration while ensuring 
the average latency/energy constraints 
in the online environment. 
This renders intractability of conventional optimization approaches. 
To tackle these challenges, we develop a low-complexity learning-driven protocol 
leveraging self-adaptive model distillation and constrained RL method in the following sections.

\section{Low-Complexity Protocol for Dynamic MEG Acceleration}
In this section, we present a low-complexity learning-driven protocol to achieve dynamic MEG acceleration.  
Note that it is extremely challenging to jointly train billions of sub-model parameters while optimizing the denoising steps and feature merging ratio in $\mathcal{P}_{0}$. 
Therefore, we decouple the customization of sub-models and the optimization of dynamic acceleration into two consecutive stages, i.e., offline distillation and online prediction. 
Specifically, the offline distillation trains a backbone architecture in a centralized fashion to approximate the performance of the original GAI model,  
which adaptively support various designs of denoising steps and feature compression, as well as overcome channel noises.  
Relying on this backbone architecture, the online prediction further achieves on-demand compression and dynamic acceleration of GAI models specific to real-time channels and task features, 
which can be modelled as a constrained MDP. 
\subsection{Offline Distillation of Few-Step Diffusion and Feature Merging}\label{Sec_DynamicDiff}
\subsubsection{Few-Step Diffusion Distillation}
LDM leverages a VAE encoder to encode high-resolution images into latent space and a VAE decoder to decode it back to the original pixel space. 
Thus, the high-complexity diffusion process can be performed over the latent space. 
During training, the VAE encoder extracts latent feature $\mathbf{Z}_{t}^{\mathrm{V}}$ from the ground-truth image of task $g_{t}$. 
Then, the denoising diffuser is learned by a forward noise addition process and a reverse denoising diffusion process over $\mathbf{Z}_{t}^{\mathrm{V}}$.
In the forward diffusion, the latent feature $\mathbf{Z}_{t}^{\mathrm{V}}$ is corrupted by Gaussia noises step by step.  
At noise addition step $l$, $l\in\mathcal{L}=\{0, 2, \ldots, L_{\max}\}$, the noise-corrupted feature is given by
\begin{equation}
    \mathbf{Z}_{t}^{(l)} = \sqrt{1-\sigma_{\mathrm{diff}}^2(l)} \mathbf{Z}_{t}^{(0)} + \sigma_{\mathrm{diff}}(l)\bm{\epsilon},
\end{equation}
where $\mathbf{Z}_{t}^{(0)}=\mathbf{Z}_{t}^{\mathrm{V}}$ is sampled from the target feature distribution, and 
$\bm{\epsilon}$ and $\sigma_{\mathrm{diff}}^2$ denote the random Gaussian noises and the variance, respectively. 
Then, during the reverse denoising diffusion process, the denoising diffuser 
learns to recover $\mathbf{Z}_{t}^{(0)}$ from $\mathbf{Z}_{t}^{(L_{\max})}$ by repeatedly executing U-Net in denoising steps $L_{\max},L_{\max}-1,...,0$. 
To achieve fast denoising diffusion, denoising diffusion implicit model (DDIM) \cite{DDIM} has been proposed to efficiently solve the probability flow ordinary differential equations (ODEs) 
based on their semi-linear structures. 
Nevertheless, DDIM still requires tens of denoising steps to maintain the generation quality, 
which are time-consuming for mobile generation. 
To overcome this limitation, we distill the high-cost denoising diffusion process into few-step denoising diffusion through the offline distillation. 
To train a backbone denoising diffuser that is applicable to variable denoising steps, 
we identify an effective distillation technique, namely rectified flow \cite{PeRFlow}. 
Rectified flow utilizes linear interpolation to reconstruct the target distribution $\varphi_{0}$ of 
desired feature $\mathbf{Z}_{t}^{(0)}$ from the noise distribution $\varphi_{l}$ of corrupted feature $\mathbf{Z}_{t}^{(l)}$ at each step $l$. 
The few-step diffusion and the original ODE solver are known as student model and teacher model, respectively.  
Given denoising steps $L$, we create $L$ time windows, where the $l$-th time window $[\tau_{l}, \tau_{l-1})$ is given by a starting point $\tau_{l}$ and an end point $\tau_{l-1}$, respectively. 
The student diffusion model learns a noise prediction neural network $\bm{\epsilon}_{\varpi}\left(\mathbf{Z}_{t}^{(\tau)}, \tau\right)$, 
parameterized by $\bm{\varpi}$, to predict noise at time point $\tau\in[\tau_{l}, \tau_{l-1})$ in each time window $l$.   
Thus, it can transport the noise-corrupted feature $\mathbf{Z}_{t}^{(l)}\sim \varphi_l$ 
into the target feature $\mathbf{Z}_{t}^{(l-1)}\sim\varphi_{l-1}$.
Specifically, the denoised feature $\mathbf{Z}_{t}^{(l-1)}$ can be estimated by 
$\bm{\epsilon}_{\varpi}\left(\mathbf{Z}_{t}^{(l)}, \tau_{l} \right)$ using
\begin{equation}\label{denoising}
    \vspace{-0.5em}
    \mathbf{Z}_{t}^{(l-1)} = u_{l} \mathbf{Z}_{t}^{(l)} 
    + v_{l} \bm{\epsilon}_{\varpi}\left(\mathbf{Z}_{t}^{(l)}, \tau_{l}\right),
    \vspace{-0.5em}
\end{equation}
where parameters $u_{l}$ and $v_{l}$ balance between current states and estimated noises. 
Following DDIM solver, we have  $u_{l} = \sqrt{\Omega_{l-1}/\Omega_{l}}$ and 
$v_{l} = \sqrt{1-\Omega_{l-1}}-\sqrt{\Omega_{l-1}\left(1-\Omega_{l}\right)/\Omega_{l}}$, 
where $\Omega_{l}=1-\sigma_{\mathrm{diff}}^{2}(l)$, $\forall l \in \mathcal{L}$. 
The student diffusion model $\bm{\epsilon}_{\varpi}\left(\mathbf{Z}_{t}^{(l)}, \tau\right)$ is trained to 
fit the linear interpolation between 
$\mathbf{Z}_{t}^{(l)}$ and $\mathbf{Z}_{t}^{(l-1)}$, $\forall l \in \mathcal{L}$.  
This results in a noise matching problem between teacher and student models, which yields the following empirical reconstruction loss for learning parameters $\bm{\varpi}$ \cite{PeRFlow}:
\begin{equation}\label{reconstruction_loss_v}
    \vspace{-0.5em}
    \mathcal{L}_{\varpi} = \sum_{l\in\mathcal{L}}\left\| \bm{\epsilon}_{\varpi}\left(\mathbf{Z}_{t}^{(\tau)},\tau\right) - \frac{\mathbf{Z}_{t}^{(l-1)} - u_{l} \mathbf{Z}_{t}^{(l)}}{v_{l}}\right\|^2, 
\vspace{-0.5em}
\end{equation}
where ${\mathbf{Z}}_{t}^{(l-1)}$ is the target feature constructed by the ODE solver $\Phi\left(\mathbf{Z}_{t}^{(l)}, \tau_{l}, \tau_{l-1}\right)$,
and $\mathbf{Z}_{t}^{(\tau)} = \frac{\tau-\tau_{l-1}}{\tau_{l}-\tau_{l-1}}\mathbf{Z}_{t}^{(l)} + \frac{\tau_{l}-\tau}{\tau_{l}-\tau_{l-1}}\mathbf{Z}_{t}^{(l-1)}$ denotes the denoised feature at time point $\tau$. 

During offline distillation, we randomly sample the number of denoising steps $L_{t}$ 
and the time point $\tau\in(\tau_{l-1},\tau_{l}]$.
Then, the few-step denoising diffusion model is trained by minimizing the reconstruction loss \eqref{reconstruction_loss_v} for noise matching. 
The denoised features, $\mathbf{Z}_{t}^{(L)}, \mathbf{Z}_{t}^{(L-1)}, ..., \mathbf{Z}_{t}^{(0)}$, can be recursively obtained using \eqref{denoising}. 
Therefore, the latent features extracted by ES-side sub-model $\mathcal{F}^{\mathrm{E}}(\cdot)$ in time frame $t$
can be given by $\mathbf{Z}_{t}^{\mathrm{E}} = \mathcal{F}^{\mathrm{E}}(\alpha_{t},\mathbf{X}_{t})$.

\subsubsection{Feature Merging for Length-Adaptive Compression} 
As the feature transmissions are crucial for collaborative generation at ES and UE, we propose a length-adaptive 
feature compression scheme in this part, 
which reduces the dimension of the transmitted feature by merging similar neurons specific to different latent features $\mathbf{Z}_{t}^{\mathrm{E}}$. 
It is worth noting that conventional pruning schemes inevitably incurs performance loss, since they directly discard the information of pruned neurons that are considered as less important. 
In comparison, we fuse the information of pruned features into the remaining neurons, thus enabling more efficient compression.  

The length-adaptive feature compression is realized by dynamically merging the neurons that are output by the last U-Net blocks.
These neurons are also known as tokens in the realms of large language model (LLM) and GAI, which is the most fundamental building blocks of Transformer. 
The idea of token merging has been recently proposed for Transformer or self-attention layer \cite{Token_Merging,Token_Merging_SD}. 
However, current studies mainly focus on computation acceleration and the token merging ratio is manually predefined. 
To support length-adaptive feature compression, our proposed protocol involves a novel dynamic feature merging scheme,   
which flexibly reduces the neurons of transmitted features in the online radio environment. 
To realize this, we first introduce an offline distillation process for learning a backbone decoder at the UE. 
Given the randomly sampled feature merging ratios, the VAE decoder at the UE is trained to reconstruct high-quality images using the merged features received through noisy wireless channels. 

Specifically, each U-Net block consists of multiple self-attention Transformer layers and convolutional layers. 
The self-attention matrix $\mathbf{A}_{t}\in\mathbb{R}^{d^{\mathrm{C}}\times d^{V}}$ of the Transformer layer in the last U-Net block is written as
\begin{equation}
    \mathbf{A}_{t}=\mathrm{softmax}\left(\frac{1}{\sqrt{d_{K}}}\mathbf{X}_{t}^{U}\mathbf{W}^{Q}\left(\mathbf{X}_{t}^{U}\mathbf{W}^{K}\right)^{T}\right)\mathbf{X}_{t}^{U}\mathbf{W}^{V},
\end{equation}
where $\mathbf{X}_{t}^{U}\in\mathbb{R}^{ d^{\mathrm{C}}\times d^{\mathrm{U}}}$ denotes the input tokens of the last U-Net block, with $d^{\mathrm{C}}$
being the number of latent feature channels, and $d^{\mathrm{U}}$ denotes the numbers of input tokens. 
Moreover, $\mathbf{W}^{K}\in\mathbb{R}^{d^{\mathrm{U}}\times d^{K}}$ 
and $\mathbf{W}^{Q}\in\mathbb{R}^{d^{\mathrm{U}} \times d^{K}}$ denote the model weights for calculating key matrix and query matrix, 
and $\mathbf{W}^{V}\in\mathbb{R}^{d^{\mathrm{U}}\times d^{V}}$ signifies the model weights for calculating value matrix. 
Thus, the unmerged latent vector $\mathbf{Z}_{t}^{\mathrm{E}}$ can be calculated by 
\begin{equation}
\mathbf{Z}_{t}^{\mathrm{E}}=\mathbf{A}_{t}\ast\mathbf{W}^{\mathrm{CONV}}\in\mathbb{R}^{d^{\mathrm{C}} \times d^{\mathrm{W}}d^{\mathrm{H}}},
\end{equation}
where $\mathbf{W}^{\mathrm{CONV}}$ denotes the 2D convolutional neural layer weights and $\ast$ is the convolutional operation. 
$d^{\mathrm{W}}$ and $d^{\mathrm{H}}$ indicate the width and height of latent features, and thus the maximum number of tokens for transmitted features can be given by 
$J_{\max} = d^{\mathrm{W}}d^{\mathrm{H}}$.

To compress latent features targets, $J_{t}=J_{\max} - \lfloor \beta_{t}J_{\max} \rceil$ useful tokens will be retained after merging.  
To determine which tokens need to be pruned and merged when the merging ratio $\beta_{t}$ is given, we measure the similarity between different tokens, 
thus merging the output tokens tailored to the time-varying latent features and channel conditions.  
It is worth noting that the token merging needs to be performed via a fast procedure, which makes conventional iterative clustering method (e.g., K-means) computationally impractical. 
To this end, we consider a low-complexity solution here. 
Specifically, we randomly divide tokens into two subsets $\mathcal{J}_{0}$ and $\mathcal{J}_{1}$. 
The cosine similarity $\Lambda_{j,j'}$ for any two tokens $j\in\mathcal{J}_{0}$ and $j'\in\mathcal{J}_{1}$ can be calculated as
\begin{equation}\label{similarity1}
    \Lambda_{j,j',t}= \frac{\mathbf{z}_{j,t}^{\mathrm{E}}\cdot\mathbf{z}_{j',t}^{\mathrm{E}}}{\left\|\mathbf{z}_{j,t}^{\mathrm{E}}\right\| \left\|\mathbf{z}_{j',t}^{\mathrm{E}}\right\|}, ~\forall m\in\mathcal{J}_{0}, ~ \forall m'\in\mathcal{J}_{1}.
\end{equation}
Then, we merge the most similar $M_{t}= J_{\max} - J_{t}$ pairs of tokens. 
Let $\Psi_{j,j',t}$ denote the binary indicator that denotes whether token $j'$ will be merged into token $j$, i.e., 
\begin{equation}\label{similarity2}
    \Psi_{j,j',t} = \begin{cases}
        1, & \text{$\Lambda_{j,j',t}$ is the top-$M_{t}$ similarity score}, \\
        0, & \text{otherwise,}
    \end{cases}
    ~\forall j\in\mathcal{J}_{0}, ~ \forall j'\in\mathcal{J}_{1}.
\end{equation}
As each merging operation reduces one token from $\mathcal{J}_{1}$, we can finally meet the required ratio and reserve $J_{t}$ tokens in $\mathcal{J}_{t}=\mathcal{J}_{0}\cup\mathcal{J}_{1}$ with lower similarities. 
By merging the similar tokens, the reserved token $j\in\mathcal{J}_{t}$ can be updated by 
\begin{equation}\label{feature_merging}
    \mathbf{z}_{j,t} = \frac{1}{\overline{\Psi}_{j,t}+1} \left(\mathbf{z}_{j,t}^{\mathrm{E}} + \sum_{j'\in\mathcal{J}_{1}}\Psi_{j,j',t}\mathbf{z}_{j',t}^{\mathrm{E}} \right), 
\end{equation}
where $\overline{\Psi}_{j,t} = \sum_{j'}\Psi_{j,j',t}$. 
Based on the above merging operation $\mathcal{F}^{\mathrm{M}}(\cdot)$, we achieve the length-adaptive latent feature: 
\begin{equation}\label{F_encoder}
    \vspace{-0.3em}
    \mathbf{Z}_{t}=\mathcal{F}^{\mathrm{M}}\left(\beta_{t}, \mathbf{Z}_{t}^{\mathrm{E}}\right)\in\mathbb{R}^{d^{\mathrm{C}}\times J_{t}}. 
    \vspace{-0.3em}
\end{equation}

The UE then receives the noisy latent feature through the wireless channels, i.e., 
\begin{equation}
    \vspace{-0.3em}
    \widehat{\mathbf{Z}}_{t} = \mathbf{Z}_{t} + \mathbf{N}_{t},
    \vspace{-0.3em}
\end{equation}
with Gaussian noises $\mathbf{N}_{t} \sim \mathcal{N}\left(0, \sigma^2\right)$. 
To keep the consistency of the feature decoder inputs, the UE further reconstructs the unmerged features $\widehat{\mathbf{Z}}_{t}^{\mathrm{U}}\in\mathbb{R}^{d^{\mathrm{C}}\times J_{\max}}$ 
by duplicating $\widehat{\mathbf{z}}_{j,t}$ to pad $\widehat{\mathbf{z}}_{j',t}$, $\forall \Psi_{j,j',t}=1$. 
Then, UE learns to decode the target images from the received noisy features by
\begin{equation}
    \vspace{-0.3em}
    \mathbf{Y}_{t} = \mathcal{F}^{\mathrm{D}}\left(\mathbf{Z}_{t}, \mathbf{N}_{t}\right).
    \vspace{-0.3em}
\end{equation}
By randomly sampling the denoising steps and feature merging ratios during the offline distillation stage, 
the VAE decoder parameters $\bm{\phi}$ can be trained to minimize the mean square error (MSE) distortion between the reference image $\mathbf{Y}_{t}^{*}$ and the reconstructed image, i.e.,  
\begin{equation}
    \mathcal{L}_{\phi} =  \sum_{t\in\mathcal{T}} \left\| \mathbf{Y}_{t}^{*}-  \mathcal{F}^{\mathrm{D}}\left(\mathbf{Z}_{t}, \mathbf{N}_{t}\right)\right\|^2, 
\end{equation}
A backbone architecture is then constructed by offline distillation, which will be utilized for dynamic compression in the online stage.

\subsection{Online MEG Acceleration: A Constrained MDP}
Based on the backbone architecture trained via offline stage, 
the online dynamic acceleration can be further formulated as a constrained MDP model $\mathcal{M}_{\mathrm{C}}$, 
which is defined as 
\begin{equation}
    \mathcal{M}_{\mathrm{C}} = \big(\mathbf{s}_{t},\mathbf{a}_{t}, \pi_{\theta}\left(\mathbf{a}_{t}\mid \mathbf{s}_{t}\right),  p\left(\mathbf{s}_{t+1}\mid \mathbf{s}_{t},\mathbf{a}_{t}\right),
\gamma,R_{t}, \mathbf{C}_{t}\big).
\end{equation} 
Specifically, at each time frame $t$, the dynamic compression predictor 
observes the environment states $\mathbf{s}_{t} = \big[h_{t},D_{t-1},E_{t-1} \big]$, 
which stacks the vectorized information of instantaneous CSI, the end-to-end MEG latency, and the system's energy consumption in the previous time frame. 
The dynamic feature merging ratios and denoising steps can be predicted by the policy $\pi_{{\theta}}$, which is parameterized by $\bm{\theta}$.
According to observations at each time frame $t$, 
the dynamic compression predictor samples the action from the policy, i.e.,
$\mathbf{a}_{t} = \left[\alpha_{t},\beta_{t}\right] \sim \pi_{\theta}\left(\mathbf{a}_{t}\mid \mathbf{s}_{t}\right)$. 
Then, the length-adaptive features $\mathbf{Z}_{t}$ can be further extracted by feature merging $\mathcal{F}^{\mathrm{M}}\left(\cdot\right)$ using \eqref{F_encoder}. 
As a result, the generated content $\mathbf{Y}_{t}$ is jointly inferred by the learning-based policy $\pi_{\theta}\left(\mathbf{a}_{t}\mid \mathbf{s}_{t}\right)$, 
the feature merging $\mathcal{F}^{\mathrm{M}}\left(\cdot\right)$, 
the feature encoder $\mathcal{F}^{\mathrm{E}}\left(\cdot\right)$, and the feature decoder $\mathcal{F}^{\mathrm{D}}\left(\cdot\right)$, i.e.,  
$\mathbf{Y}_{t} = 
\mathcal{F}^{\mathrm{D}}\left(\mathcal{F}^{\mathrm{M}}\left(\beta_{t},\mathcal{F}^{\mathrm{E}}\left(\alpha_{t},\mathbf{X}_{t}\right)\right), \mathbf{N}_{t}\right)$.
In addition, the reward function $R_{t} = F_{\mathrm{G}}\left(\mathbf{Y}_{t},\mathbf{a}_{t}\mid\mathbf{s}_{t}\right)$ 
indicates the image generation qualities measured by the environment feedback. 
$\mathbf{C}_{t} = [C_{t,1}\left(\mathbf{s}_{t},\mathbf{a}_{t}\right), C_{t,2}\left(\mathbf{s}_{t},\mathbf{a}_{t}\right)]$ represent the cost functions corresponding to latency and energy consumption constraint violations, 
where $C_{t,1}\left(\mathbf{s}_{t},\mathbf{a}_{t}\right)=\left[D_{t}\left(\alpha_{t},\beta_{t}\right)-D_{\max}\right]^{+}$ 
and $C_{t,2}\left(\mathbf{s}_{t},\mathbf{a}_{t}\right)=\left[E_{t}\left(\alpha_{t},\beta_{t}\right)-E_{\max}\right]^{+}$, respectively. 
The state transitions from $\mathbf{s}_{t}$ to $\mathbf{s}_{t+1}$ under the action $\mathbf{a}_{t}$ can be specified by the state transition probability $p\left(\mathbf{s}_{t+1}\mid \mathbf{s}_{t},\mathbf{a}_{t}\right)$,  
which leads to a trajectory $\rho = \left[\mathbf{s}_{1},\mathbf{a}_{1}, R_{1},  \dots, \mathbf{s}_{T},  \mathbf{a}_{T}, R_{T}\right]$. 

To sum up, our goal is to maximize the discounted return 
$G^{\mathrm{R}}\left(\pi_{\theta}\left(\mathbf{a}\mid \mathbf{s}\right)\right)\!\!=\!\!\!\!\mathop{\mathbb{E}}
\limits_{\rho\sim\pi_{\theta}\left(\mathbf{a}\mid \mathbf{s}\right)}\!\!\left[\sum\limits_{t=0}^{\infty}\gamma^{t}R\left(\mathbf{s}_{t},\mathbf{a}_{t}\right)\right]$ with discount factor $\gamma \in [0,1]$, 
subject to the violations of the expected discounted costs of latency constraint 
$G_{1}^{\mathrm{C}}\left(\pi_{\theta}\left(\mathbf{a}\mid \mathbf{s}\right)\right)
=\mathop{\mathbb{E}}\limits_{\rho\sim \pi_{\theta}\left(\cdot|\mathbf{s}\right)}
\left[\sum\limits_{t=0}^{\infty}\gamma^{t}C_{t,1}\left(\mathbf{s}_{t},\mathbf{a}_{t}\right)\right]$ 
and energy constraint $G_{2}^{\mathrm{C}}\left(\pi_{\theta}\left(\mathbf{a}\mid \mathbf{s}\right)\right)
=\mathop{\mathbb{E}}\limits_{\rho\sim \pi_{\theta}\left(\cdot|\mathbf{s}\right)}
\left[\sum\limits_{t=0}^{\infty}\gamma^{t}C_{t,2}\left(\mathbf{s}_{t},\mathbf{a}_{t}\right)\right]$  
up to prescribed thresholds $\varepsilon_{i} \in [0,+\infty]$, $i \in \{0,1\}$, which is formulated as
\begin{subequations}\label{P_CMDP}
    \begin{align*}
    \mathcal{P}_{\mathrm{CMDP}}: &  \max_{\pi_{\theta}\left(\mathbf{a}\mid \mathbf{s}\right)} ~ G^{\mathrm{R}} \left(\pi_{\theta}\left(\mathbf{a}\mid \mathbf{s}\right)\right) \tag{\ref{P_CMDP}{a}}
    \\ {\mathrm{s.t.}} ~ & \mathrm{C}3, \tag{\ref{P_CMDP}{b}}
    \\ & \mathrm{C}4: ~ G_{i}^{\mathrm{C}}\left(\pi_{\theta}\left(\mathbf{a}\mid \mathbf{s}\right)\right)\leqslant \varepsilon_{i}, ~ \forall i \in \{1,2\}. \tag{\ref{P_CMDP}{c}}    \end{align*}
\end{subequations}

\section{Constrained Reinforcement Learning for Online MEG Acceleration}
In this section, we solve the constrained MDP formulated in the previous section for online MEG acceleration. 
To achieve higher image quality while guaranteeing both latency and energy constraints, 
we develop the MEG-CVPO algorithm based on the constrained RL.

\subsection{Preliminaries on Constrained RL Method}
We commence with revisiting the conventional Lagrangian-based method \cite{Lagrangian}, 
and then introduce the CVPO theory \cite{CVPO} to achieve more efficient constraint guarantees for online prediction.
\subsubsection{Conventional Lagrangian-based Method}
We first introduce the conventional Lagrangian-based method that is commonly adopted in solving constrained MDP. 
The main principle is to transfer the constrained MDP problem $\mathcal{P}_{\mathrm{CMDP}}$ into a min-max optimization problem:
\begin{equation}\label{P_Larange}
    \left(\bm{\theta}^{*},\bm{\lambda}^{*}\right)=\arg\min_{\bm{\lambda}\geqslant0}
    \max\limits_{\bm{\theta}}~G^{\mathrm{R}}\left(\pi_{\theta}\left(\mathbf{a}\mid \mathbf{s}\right)\right)
    -\sum_{i\in\{1,2\}}\lambda_{i}\left(G_{i}^{\mathrm{C}}\left(\pi_{\theta}\left(\mathbf{a}\mid \mathbf{s}\right)\right)-\varepsilon_{i}\right), 
\end{equation}
where $\lambda_{i}$, $i = 1,2$,  is the non-negative Lagrange multiplier (i.e., dual variable) corresponding to constraint $\mathrm{C}4$.
The constrained MDP can be addressed by alternatively improving policy parameters $\bm{\theta}$ and updating Lagrange multipliers $\bm{\lambda}=\left[\lambda_{1},\lambda_{2}\right]$ to iteratively solve the min-max problem. 
Note that the optimal dual variables are expected to satisfy $\lambda_{i}^{\ast} \rightarrow +\infty$ when $G_{i}^{\mathrm{C}}>\varepsilon_{i}$ 
and $\lambda_{i}^{\ast}=0$ when $G_{i}^{\mathrm{C}}\leqslant \varepsilon_{i}$, $\forall i \in \{1,2\}$. 
However, due to the stochastic nature of channel environment and thus the cost penalty terms, the primal problem is hard to optimize via the policy gradients backpropagated from multiple Q-value functions. 

\subsubsection{Variational Policy Optimization Method}
To overcome the above issues, the constrained RL for dynamic MEG acceleration can be transformed into a probabilistic variational inference problem, which is elaborated on in the sequel. 
 
First, we define a variable ${O}$ to represent the event of obtaining the optimal RL reward maximization. 
Given a trajectory $\rho$, the likelihood of RL optimality is proportional to the exponential cumulative discount rewards \cite{CVPO,RL_optimality}, which can be written as
\begin{equation}\label{Prob_O1}
    p\left(O=1\mid\rho\right)\propto\exp\left(\frac{1}{\kappa}\sum_{t=0}^{\infty}\gamma^{t}R_{t}\right),
    \end{equation}
where $\kappa$ denotes the temperature parameter. 
The log-likelihood of RL optimality under the policy $\pi_{\theta}\left(\mathbf{a}_{t}\mid \mathbf{s}_{t}\right)$, 
denoted by $\log \left(p_{\pi}\left(O=1\right) \right)$, has the following lower bound $U\left(q,\pi\right)$: 
\begin{equation}\label{Log_Prob_O1}
    \begin{split}
     \log \left(p_{\pi}\left(O=1\right) \right)
     & =\log\left(\int p\left(O=1\mid \rho\right)p_{\pi}\left(\rho\right)d\rho  \right)
    = \log \mathbb{E}_{\rho\sim q\left(\mathbf{a}\mid \mathbf{s}\right)} \left[\frac{p(O=1\mid \rho)p_{\pi}(\rho)}{p_{q}\left(\rho\right)}\right] \\
    & \overset{(a)}{\geqslant}  \mathbb{E}_{\rho\sim q\left(\mathbf{a}\mid \mathbf{s}\right)} \left[\log\frac{p\left(O=1\mid \rho\right)p_{\pi}\left(\rho\right)}{p_{q}\left(\rho\right)}\right] \\&
    = \mathbb{E}_{\rho\sim q\left(\mathbf{a}\mid \mathbf{s}\right)}\left[\log p\left(O=1\mid \rho\right)\right] 
    - D_{\mathrm{KL}}\left(p_{q}\left(\rho\right)\|p_{\pi}\left(\rho\right)\right)
    \triangleq U\left(q,\pi\right), 
    \end{split}
\end{equation} 
where inequality (a) results from Jensen inequality.  
$p_{\pi}\left(\rho\right)$ and $p_{q}\left(\rho\right)$ denote the probabilities of obtaining trajectory $\rho$ under policy $\pi_{\theta}\left(\mathbf{a}_{t}\mid \mathbf{s}_{t}\right)$ 
and auxiliary policy $q\left(\mathbf{a}_{t}\mid \mathbf{s}_{t}\right)$, respectively.  
Moreover, $D_{\mathrm{KL}}\left(p_{q}\left(\rho\right)\|p_{\pi}\left(\rho\right)\right)$ 
is the Kullback-Leibler (KL) divergence between distributions 
$p_{q}\left(\rho\right)$ and $p_{\pi}\left(\rho\right)$, which is defined as 
\begin{equation}
    D_{\mathrm{KL}}\left(p_{q}\left(\rho\right)\|p_{\pi}\left(\rho\right)\right) 
    \triangleq \mathbb{E}_{\rho\sim q\left(\mathbf{a}\mid \mathbf{s}\right)}\left[\log\frac{p_{q}\left(\rho\right)}{p_{\pi}(\rho)}\right]
    = - \mathbb{E}_{\rho\sim q\left(\mathbf{a}\mid \mathbf{s}\right)}\left[\log\frac{p_{\pi}(\rho)}{p_{q}\left(\rho\right)}\right]. 
\end{equation}
Substituting \eqref{Prob_O1} into \eqref{Log_Prob_O1} and multiplying with $\kappa$ 
leads to the following evidence lower bound (ELBO), denoted by $U_{\mathrm{E}}\left(q,\pi\right)$, 
for the log-likelihood of RL optimality: 
\begin{equation}\label{ELBO}
    U\left(q, \pi\right)
    \propto  \mathbb{E}_{\rho\sim q\left(\mathbf{a}\mid \mathbf{s}\right)}\left[\sum_{t=0}^{\infty}\gamma^{t}R_{t}\right]
    -\kappa D_{\mathrm{KL}}\left(p_{q}\left(\rho\right)\|p_{\pi}\left(\rho\right)\right) 
    \triangleq  U_{\mathrm{E}}\left(q,\pi\right),
    \end{equation}
The auxiliary policy $q\left(\mathbf{a}\mid \mathbf{s}_{t}\right)$ is introduced to 
approximate $\pi_{\theta}\left(\mathbf{a}_{t}\mid \mathbf{s}_{t}\right)$ while satisfying constraints. 
More specifically, $p_{\pi}(\rho)$ and $p_{q}\left(\rho\right)$ can be expressed by
\begin{subequations}\label{action_distribution}
    \begin{equation}
        p_{\pi}\left(\rho\right)=p\left(\mathbf{s}_{0}\right)\prod_{t\geqslant0}p\left(\mathbf{s}_{t+1}\mid\mathbf{s}_{t},\mathbf{a}_{t}\right)\pi\left(\mathbf{a}_{t}\mid\mathbf{s}_{t}\right)p\left(\bm{\theta}\right),
    \end{equation}
    \begin{equation}
    p_{q}\left(\rho\right)=p\left(\mathbf{s}_{0}\right)\prod_{t\geqslant0}p\left(\mathbf{s}_{t+1}\mid\mathbf{s}_{t},\mathbf{a}_{t}\right)q\left(\mathbf{a}_{t}\mid\mathbf{s}_{t}\right),
    ~ \forall q\left(\mathbf{a}\mid \mathbf{s}\right)\in\Pi_{q}, 
    \end{equation}
\end{subequations}
where  $\Pi_{q}$  represents a set of feasible policy distribution subject to system constraints, which is given by
\begin{equation}
    \Pi_{q}=\big\{ q\left(\mathbf{a}\mid\mathbf{s}\right):~G_{i}^{\mathrm{C}}\left(q\left(\mathbf{a}\mid\mathbf{s}\right)\right) \leqslant \varepsilon_{i},
    ~ \forall i\in\{1,2\}, ~ \mathbf{a}\in\mathcal{A}, ~\mathbf{s}\in\mathcal{S}\big\}.
    \end{equation}
Next, substituting \eqref{action_distribution} for $p_\pi(\rho)$ and  $p_{q}\left(\rho\right)$ into \eqref{ELBO}, we can obtain
\begin{equation}\label{Lower_Bound}
    U_{\mathrm{E}}\left(q,\pi\right)\!
    =\!{\mathbb{E}}_{\rho\sim q\left(\mathbf{a}\mid \mathbf{s}\right)}\left[\sum_{t=0}^{\infty}\gamma^{t}R_{t}\right]
    -\!\kappa D_{\mathrm{KL}}\left(q\left(\mathbf{a}\mid\mathbf{s}\right) \| \pi_{\theta}\left(\mathbf{a}\mid\mathbf{s}\right)\right)
     +\kappa\log p\left(\bm{\theta}\right),  
     ~ q\left(\mathbf{a}\mid\mathbf{s}\right)\in\Pi_{q}. 
    \end{equation}

CVPO aims for solving problem \eqref{P_CMDP} by alternating between the Expectation step (E-step) and the Maximization step (M-step) for constrained policy learning.
Specifically, the E-step optimizes the lower bound $U\left(q,\pi\right)$ with respect to $q\left(\mathbf{a}\mid \mathbf{s}\right)$ 
within the feasible policy distribution set $\Pi_{q}$; 
and the M-step optimizes parameters $\bm{\theta}$ of policy $\pi_{\theta}\left(\mathbf{a}\mid \mathbf{s}\right)$.
The above expectation-maximization (EM) procedure decouples the variational distribution approximation of feasible policies and the policy improvement 
leveraging an auxiliary policy specified  by $q\left(\mathbf{a}\mid \mathbf{s}\right)\in\Pi_q$.  

\subsection{The Proposed MEG-CVPO Algorithm}
Based on the CVPO theory, the main difficulty of solving problem \eqref{P_CMDP} turns into searching for the variational distribution within the constrained set. 
Fortunately, by treating $q\left(\mathbf{a}\mid \mathbf{s}\right)$ as variables (instead of parameterized neural networks), the constrained distribution $q\left(\mathbf{a}\mid \mathbf{s}\right)$ can be solved analytically by convex optimization to guarantee both the optimality and the feasibility.
In the following part, we will develop the constrained MEG policy learning algorithm by combining the convex optimization  and the supervised learning.
\subsubsection{E-Step Optimization} 
In each training iteration $n$, the E-step aims to search for the optimal variational distribution $q\left(\mathbf{a}\mid \mathbf{s}\right)\in\Pi_{q}$, which guarantees the constraints whilst improving expected returns. 
Given the sampled trajectory $\rho=\left\{\left(\mathbf{s}_{t},\mathbf{a}_{t},R_{t}\right)\right\}_{t=1}^{T}$ from the replay buffer, the goal is to optimize the following KL objective  obtained from \eqref{Lower_Bound}:
\begin{subequations}\label{P_E_step}
    \begin{align*}
    \mathcal{P}_{\mathrm{E}}: \max_{q\left(\mathbf{a}\mid \mathbf{s}\right)} ~& 
    U_{\mathrm{E}}\left(q,\pi\right)={\mathbb{E}}_{\mathbf{s}\sim\psi}\left[\mathbb{E}_{\rho\sim q\left(\mathbf{a}\mid \mathbf{s}\right)}\left[Q_{n}^{\mathrm{R}}\left(\mathbf{s},\mathbf{a}\right)\right]
    -\kappa D_{\mathrm{KL}}\left(q\|\pi_{n}\right)\right] \tag{\ref{P_E_step}{a}}
    \\ {\mathrm{s.t.}}~ &
    \mathrm{C}3, \tag{\ref{P_E_step}{b}}
    \\ &
    \mathrm{C}5: ~ {\mathbb{E}}_{\mathbf{s}\sim\psi}\left[\mathbb{E}_{\rho\sim q\left(\mathbf{a}\mid \mathbf{s}\right)}\left[Q_{i,n}^{\mathrm{C}}\left(\mathbf{s},\mathbf{a}\right)\right]\right]\leqslant\varepsilon_{i}, 
    ~ \forall i \in \{1,2\}, \label{constraint_E_step} \tag{\ref{P_E_step}{c}}
    \end{align*}
\end{subequations}
where constraint $\mathrm{C}$4 is rewritten into  $\mathrm{C}$5, 
$\psi$ denotes the stationary state distribution based on current variational distribution, 
and the reward value $Q_{n}^{\mathrm{R}}\left(\mathbf{s},\mathbf{a}\right)$
and the cost value $Q_{i,n}^{\mathrm{C}}\left(\mathbf{s},\mathbf{a}\right)$ 
are defined as 
\vspace{-0.3em}
\begin{equation}
    Q_{n}^{\mathrm{R}}\left(\mathbf{s},\mathbf{a}\right)
    =\mathbb{E}_{\rho\sim\pi_{n}\left(\mathbf{a}\mid \mathbf{s}\right),\atop\mathbf{s}_{0}=\mathbf{s},\mathbf{a}_{0}=\mathbf{a}}\left[\sum\limits_{t=0}^{\infty}\gamma^{t}R_{t}\left(\mathbf{s}_{t},\mathbf{a}_{t}\right)\right],
    \quad
    Q_{i,n}^{\mathrm{C}}\left(\mathbf{s},\mathbf{a}\right)={\mathbb{E}}_{\rho\sim\pi_{n}\left(\mathbf{a}\mid \mathbf{s}\right),\atop\mathbf{s}_{0}=\mathbf{s},\mathbf{a}_{0}=\mathbf{a}}\left[\sum_{t=0}^{\infty}\gamma^{t}C_{t,i}\left(\mathbf{s}_{t},\mathbf{a}_{t}\right)\right]. 
    \vspace{-0.3em}
\end{equation}
Here, $Q_{n}^{\mathrm{R}}\left(\mathbf{s},\mathbf{a}\right)$ and $Q_{i,n}^{\mathrm{C}}\left(\mathbf{s},\mathbf{a}\right)$ 
are estimated over the sampled trajectory $\rho\sim\pi_{n}$.
Constraint $\mathrm{C}5$ guarantees that the optimized variational distribution $q$ lies within the feasible set $\Pi_{q}$. 
Problem $\mathcal{P}_{\mathrm{E}}$ can be viewed as a constrained KL-regularized optimization problem. 
To avoid tuning the penalty coefficient $\kappa$ in the regularized term, we can further introduce a hard regularization constraint as follows
\begin{subequations}\label{P_E_Hard}
    \vspace{-0.3em}
    \begin{align*}
    \widetilde{\mathcal{P}}_{\mathrm{E}}: & \max_{q\left(\mathbf{a}\mid\mathbf{s}\right)} ~ 
    {\mathbb{E}}_{\mathbf{s}\sim\psi}\left[\int q\left(\mathbf{a}\mid\mathbf{s}\right)Q_{n}^{\mathrm{R}}\left(\mathbf{s},\mathbf{a}\right)d\mathbf{a}\right] \tag{\ref{P_E_Hard}{a}}
    \\ {\mathrm{s.t.}}~ &
    \mathrm{C}5: {\mathbb{E}}_{\mathbf{s}\sim\psi}\left[\int q\left(\mathbf{a}\mid\mathbf{s}\right)Q_{i,n}^{\mathrm{C}}\left(\mathbf{s},\mathbf{a}\right)d\mathbf{a}\right]\leqslant\varepsilon_{i}, \!
    ~ \forall i \in \{1,2\}, \label{constraint_CMDP_Reg} \tag{\ref{P_E_Hard}{b}}
    \\ &
    \mathrm{C}6:{\mathbb{E}}_{\mathbf{s}\sim\psi}\left[D_{\mathrm{KL}}\left(q\|\pi_{n}\right)\right]\leqslant {\varepsilon}_{0}, \label{constraint_Reg} \tag{\ref{P_E_Hard}{c}}
    \\ &
    \mathrm{C}7: \int q\left(\mathbf{a}\mid\mathbf{s}\right)d\mathbf{a}=1, ~ \forall \mathbf{s}\sim \psi, \label{constraint_int} \tag{\ref{P_E_Hard}{d}}
    \end{align*}
\end{subequations}
where $\mathrm{C}5$ is the cost constraint rearranged from \eqref{constraint_E_step}, 
$\mathrm{C}6$ is the regularization constraint that limits variational distribution $q$ within a trust region of 
the policy distribution at training iteration $n$, 
and $\mathrm{C}7$ ensures that variational distribution $q$ leads to valid actions across all the states. 

Regarding $q\left(\mathbf{a}\mid\mathbf{s}\right)$ as a optimization variable instead of a parameterized function, 
problem \eqref{P_E_Hard} becomes a convex optimization problem. 
The Lagrangian function of the constrained E-step optimization problem can be written as 
\vspace{-0.3em}
\begin{multline}\label{Lagrangian_Estep}
\mathcal{L}_{\lambda}\left(q\left(\mathbf{a}\mid\mathbf{s}\right),\zeta,\bm{\lambda},\xi\right)
=\!\int\!\psi\!\left(\mathbf{s}\right)\!\int\!\!q\left(\mathbf{a}\mid\mathbf{s}\right)Q_{n}^{\mathrm{R}}\left(\mathbf{s},\mathbf{a}\right)d\mathbf{a}d\mathbf{s}
+\xi\left(1\!-\!\int\!\psi\left(\mathbf{s}\right)\!\int\! q\!\left(\mathbf{a}\!\mid\mathbf{s}\right)d\mathbf{a}d\mathbf{s}\right)
\\+\sum_{i}\lambda_{i}\!\left(\!\varepsilon_{i}\!-\!\!\int\!\!\psi\!\left(\mathbf{s}\right)\!\int\!\! q\left(\mathbf{a}\!\mid\!\mathbf{s}\right)Q_{i,n}^{\mathrm{C}}\!\left(\mathbf{s},\mathbf{a}\right)\!d\mathbf{a}d\mathbf{s}\right)
\!\!+\zeta\!\left(\!\varepsilon_{0} \!-\!\!\int\!\psi\left(\mathbf{s}\right)\!\!\int\!\! q\left(\mathbf{a}\mid\mathbf{s}\right)\log\frac{q\left(\mathbf{a}\mid\mathbf{s}\right)}{\pi_{n}\left(\mathbf{a}\mid\mathbf{s}\right)}d\mathbf{a}d\mathbf{s}\right),
\vspace{-0.3em}
\end{multline}
where $\bm{\lambda}$, $\zeta$, and $\xi$ denote the Lagrangian multipliers for constraints $\mathrm{C}5 - \mathrm{C}7$. 
Then, the dual problem is given by
\vspace{-0.3em}
\begin{equation}
    \min_{\zeta,\bm{\lambda},\xi} ~ 
    \max_{q\left(\mathbf{a}\mid\mathbf{s}\right)} ~
    \mathcal{L}_{\lambda} \left(q\left(\mathbf{a}\mid\mathbf{s}\right),\zeta,\bm{\lambda},\xi\right). 
\vspace{-0.3em}
\end{equation}
The derivative of $\mathcal{L}_{\lambda}$ with respect to $q\left(\mathbf{a}\mid\mathbf{s}\right)$ can be given by
\vspace{-0.3em}
\begin{equation}
    \frac{\partial\mathcal{L}_{\lambda}}{\partial q}=Q_{n}^{\mathrm{R}}\left(\mathbf{s},\mathbf{a}\right)
    -\sum_{i\in\{1,2\}}\lambda_{i}Q_{i,n}^{\mathrm{C}}\left(\mathbf{s},\mathbf{a}\right)
    -\zeta\left(1+\log\frac{q\left(\mathbf{a}\mid\mathbf{s}\right)}{\pi_{n}\left(\mathbf{a}\mid\mathbf{s}\right)}\right)-\xi.
    \vspace{-0.3em}
\end{equation}
Based on the first-order optimality, i.e., $\frac{\partial\mathcal{L}_{\lambda}}{\partial q}=0$, 
the optimal variational distribution within $\Pi_{q}$ for problem $\widetilde{\mathcal{P}}_{\mathrm{E}}$ is given by 
\vspace{-0.3em}
\begin{equation}\label{Variational_Dist_Update}
    q^{*}\left(\mathbf{a}\mid\mathbf{s}\right)=\frac{\pi_{n}\left(\mathbf{a}\mid\mathbf{s}\right)}{B_{n}}
    \exp\left(\frac{A_{n}(\mathbf{s},\mathbf{a})}{\zeta}\right), 
    \vspace{-0.3em}
\end{equation}
where $B_{n}=\exp\left(1+{\xi}/{\zeta}\right)$ means the constant normalizer for the optimized distribution $q$, 
and $A_{n}\left(\mathbf{s},\mathbf{a}\right)$ is given by 
\begin{equation}\label{Weighted_Q}
    A_{n}\left(\mathbf{s},\mathbf{a}\right) \triangleq Q_{n}^{\mathrm{R}}\left(\mathbf{s},\mathbf{a}\right)-\sum_{i}\lambda_{i}Q_{i,n}^{\mathrm{C}}\left(\mathbf{s},\mathbf{a}\right).
    \end{equation}
Substituting \eqref{Variational_Dist_Update} into \eqref{Lagrangian_Estep} and  ignoring irrelevant terms, the optimal dual variables $\zeta^{*}$ and $\bm{\lambda}^{*}$ can be obtained by optimizing the unconstrained convex problem
\begin{equation}\label{P_E_Dual}
    \min_{\zeta,\bm{\lambda}\geqslant 0} ~ F\left(\zeta,\bm{\lambda}\right)
    = \sum_{i\in\{1,2\}}\lambda_{i}\varepsilon_{i} + \zeta\varepsilon_{0} 
    + \zeta \mathbb{E}_{\mathbf{s}\sim\psi} \left[ \log \mathbb{E}_{\mathbf{a}\sim\pi_{n}} \left[\exp\left(\frac{A_{n}\left(\mathbf{s},\mathbf{a}\right)}
    {\zeta}\right)\right]\right], 
    \end{equation}
where $F\left(\zeta,\bm{\lambda}\right)$ is the Lagrangian function.
This unconstrained convex problem can be directly solved by performing gradient descent updates over ${\zeta}$ and $\bm{\lambda}$ as follows:
\begin{equation}\label{Dual_Update}
    {\zeta} \leftarrow {\zeta} - \nu_{\zeta} \frac{\partial{F\left(\bm{\zeta},\zeta\right) }}{\partial \zeta}, 
    \quad
    \bm{\lambda} \leftarrow \bm{\lambda} - \nu_{\lambda} \frac{\partial{F\left(\zeta,\bm{\lambda}\right) }}{\partial \bm{\lambda}}, 
    \end{equation}
where $\nu_{\lambda}$ and $\nu_{\zeta}$ denote the learning rates for dual variables $\bm{\lambda}$ and $\zeta$, respectively.

\subsubsection{M-Step Optimization}
After obtaining an optimal feasible variational distribution $q_{n}^{*}$ from the E-step optimization, 
the M-step further maximizes the ELBO to update the policy $\pi_{\bm{\theta}}=\pi\left(\mathbf{a}_{t}\mid\mathbf{s}_{t}\right)$, 
where $\bm{\theta}$ denotes the learnable policy parameters.
Ignoring irrelevant terms in \eqref{Lower_Bound} with respect to $\bm{\theta}$, the M-step solves the following posterior maximization problem
\vspace{-0.3em}
\begin{equation}\label{M_Step_Objective}
    \max_{\bm{\theta}}\int\psi\left(\mathbf{s}\right)\int q^{*}\left(\mathbf{a}\mid\mathbf{s}\right)\log\pi_{\theta}\left(\mathbf{a}\mid\mathbf{s}\right)d\mathbf{a}d\mathbf{s}+\log p\left(\bm{\theta}\right), 
    \vspace{-0.3em}
\end{equation}
Regarding the policy parameters distribution $p\left(\bm{\theta}\right)$, we sample the M-step policy parameters around the old policy parameters $\bm{\theta}_{n}$ with a Gaussian prior regularizer, i.e., 
$\bm{\theta} \sim \mathcal{N}\left(\bm{\theta}_{n},\frac{F_{\theta}}{\mu}\right)$, with $\Sigma_{\theta}$ being the Fisher information matrix and $\mu$ being a positive constant coefficient. 
Based on the Gaussian prior, the log-prior can be written as 
\vspace{-0.3em}
\begin{equation}
   \log p\left(\bm{\theta}\right) = \! -\mu\left(\bm{\theta}\!-\!\bm{\theta}_{n}\right)^{T}\Sigma_{\theta}^{-1}\left(\bm{\theta}\!-\!\bm{\theta}_{n}\right) 
   \overset{(a)}{\geqslant} 
   -\mu D_{\mathrm{KL}}\left(\pi_{n}||\pi_{\theta}\right), 
   \end{equation}
where $(a)$ can be obtained by second-order Taylor expansion of $D_{\mathrm{KL}}\left(\pi_{n}\|\pi_{\bm{\theta}}\right)$. 
Therefore, the M-step objective function in \eqref{M_Step_Objective} can be transferred into maximizing the tight lower bound
\begin{equation}
    \max_{\bm{\theta}}\!\int\!\psi\!\left(\mathbf{s}\right)\!\int\!\left(q^{*}\!\left(\mathbf{a}\mid\mathbf{s}\right)\log\pi_{\bm{\theta}}\left(\mathbf{a}\mid\mathbf{s}\right)d\mathbf{a}\!-\!\mu D_{\mathrm{KL}}\left(\pi_{n}\|\pi_{\bm{\theta}}\right)\right)d\mathbf{s}.
\end{equation}

\begin{algorithm}[!tbp]
    \flushleft
    \caption{Training Procedure of MEG-CVPO Algorithm}\label{alg:MEG-CVPO}
    \begin{algorithmic}[1]
        \Require{Constrained RL batch size $B_{\mathrm{RL}}$, particle size $K$ for variational policy optimization, numbers of gradient update iterations  $I_{\mathrm{E}}$ and $I_{\mathrm{M}}$ for E-step and M-step.}
        \Ensure{Constrained policy parameters $\bm{\theta}$.}
        \State Initialize policy parameters $\bm{\theta}_{0}$.
        \For{each iteration $n=1,2,...$}
            \Statex~~~~\textit{//* Perform MEG and collect samples}
            \For{each batch execution epoch}
                \State Sample $\alpha_{t}$ and $\beta_{t}$ according to policy $\pi_{n}$ by observing $\mathbf{s}_{t}$ at each time frame $t$. 
                \State The ES extracts feature $\mathbf{Z}_{t}^{\mathrm{E}}\in\mathbb{R}^{d^{\mathrm{C}}\times J_{\max}}$ using sub-model $\mathcal{F}^{\mathrm{E}}\left(\alpha_{t},\mathbf{X}_{t}\right)$. 
                \State The ES computes similarity  by \eqref{similarity1} and performs feature merging via \eqref{similarity2} - \eqref{feature_merging}.
                \State The UE receives feature $\widehat{\mathbf{Z}}_{t}\in\mathbb{R}^{d^{\mathrm{C}}\times J_{t}}$ corrupted by Gaussian noise $\mathbf{N}_{t}$ from the ES. 
                \State The UE unmerges feature and decodes image $\mathbf{Y}_{t}$ by sub-model  $\mathcal{F}^{\mathrm{D}}(\cdot)$.
                \State Evaluate the generation quality and update replay buffer by the collected transitions.
            \EndFor
            \Statex~~~~\textit{//* Perform E-step policy optimization}
            \State Sample a set of $\mathcal{B}_{n}$ transitions.
            \State Update critic networks $Q_{n}^{\mathrm{R}}\left(\mathbf{s},\mathbf{a}\right)$ and $Q_{i,n}^{\mathrm{C}}\left(\mathbf{s},\mathbf{a}\right)$, $i \in \{1,2\}$, by Bellman backup.
            \For{each batch $b=1,2,..,B_{\mathrm{RL}}$}
                \State Sample $K$ actions $\mathbf{a}_{1}, \mathbf{a}_2,..., \mathbf{a}_{K}$ for current state $\mathbf{s}_{n}$.
                \State Calculate $Q_{n}^{\mathrm{R}}\left(\mathbf{s}_{n},\mathbf{a}_{k}\right)$ and $Q_{i,n}^{\mathrm{C}}\left(\mathbf{s}_{n},\mathbf{a}_{k}\right)$, $\forall k$.
            \EndFor
            \State Update dual variable $\zeta$ and $\bm{\lambda}$ via \eqref{Dual_Update} for $I_{\mathrm{E}}$ iterations to solve \eqref{P_E_Dual}.
            \State Update the variational distribution for state $\mathbf{s}\in\mathcal{B}_{n}$ using \eqref{Variational_Dist_Update}.
            \Statex~~~~\textit{//* Perform M-step policy optimization} 
            \State Update policy parameters $\bm{\theta}_{n}$ via \eqref{Policy_Update} for $I_{\mathrm{M}}$ iterations.
        \EndFor
    \end{algorithmic}
\end{algorithm}

Similar to E-step, 
to eliminate the hyperparameter $\mu$, 
we can transform the above soft regularized KL term into the hard KL constraint $D_{\mathrm{KL}}\left(\pi_{n}||\pi_{\bm{\theta}}\right)\leqslant \widetilde{\varepsilon}$.
This leads to the following M-step Lagrangian function
\begin{equation}
        \mathcal{L}_{\lambda}\left(\bm{\theta},\omega\right)
    =\int\psi\left(\mathbf{s}\right)\int q^{*}\left(\mathbf{a}\mid\mathbf{s}\right)\log\pi_{\bm{\theta}}\left(\mathbf{a}\mid\mathbf{s}\right)d\mathbf{a}d\mathbf{s}
     +\omega\left(\widetilde{\varepsilon}-C_{\theta}\right), 
\end{equation}
with $C_{\theta}\triangleq \left(\bm{\theta}-\bm{\theta}_{n}\right)^{T}F_{\theta}^{-1}\left(\bm{\theta}-\bm{\theta}_{n}\right)$
and $\omega$ being the dual variable for hard KL constraint.
Therefore, the dual problem is given by 
\begin{equation}\label{M_Lagrange}
    \max_{\bm{\theta}}\min_{\omega}\mathcal{L}_{\lambda}\left(\bm{\theta},\omega\right).
\end{equation}

Based on gradient descent, the policy parameters can be updated by
\begin{equation}\label{Policy_Update}
    \bm{\theta}\leftarrow\bm{\theta}+\nu_{\theta}\frac{\partial\mathcal{L}_{\lambda}\left(\bm{\theta},\omega\right)}{\partial\bm{\theta}}, 
    \quad
    \omega\leftarrow\omega-\nu_{\omega}\frac{\partial\mathcal{L}_{\lambda}\left(\bm{\theta},\omega\right)}{\partial\omega},
    \end{equation}
where $\nu_{\theta}$ and $\nu_{\omega}$ denote the learning rates. 
For ease of implementation, the integral operation is replaced by summation by sampling $K$ particles in the continuous action space.
The training procedure of MEG-CVPO algorithm can be summarized in \textbf{Algrorithm \ref{alg:MEG-CVPO}}.

\section{Numerical Results: A Case Study of \textit{SDXL} Acceleration}

\subsection{Experimental Setup}
To characterize the performance, we focus on the case studies for dynamic acceleration of \textit{Stable Diffusion XL} (\textit{SDXL}) \cite{SDXL}, which is a large-scale LDM extensively applied in a broad range of GAI applications, 
e.g., text-to-image generation, image editing/synthesis, and music generation.  
\textit{SDXL} surpasses the behaviour performance of all previous stable diffusion (SD) models with $3$x larger U-Net blocks and advanced conditioning scheme. 
While \textit{SDXL} is capable of generating high-resolution images (typically $1024\times 1024$ pixels), 
the intensive attention blocks in U-Net backbone and the increased latent feature sizes results in extraordinarily unaffordable computation burdens to directly perform it at mobile terminals. 

\begin{table}[!h]
    \centering
    \caption{Simulation parameters}
    \resizebox{1\textwidth}{!}{ 
        \begin{tabular}{l|c|l|c}
            \toprule
            Parameter & Value & Parameter & Value\\
            \hline
            Radius & $300$ [m]                                    & Latency model & $f^{\mathrm{ES}} = 1/0.0274$ [TFLOPS]\\
            Pathloss & $35.3+37.\log(dist\text{[m]})$ [dB]        & $\quad$ & $f^{\mathrm{UE}}=0.5f^{\mathrm{ES}}$ [TFLOPS]\\
            AWGN & $-94$ [dBm]                                      & \quad & $D_{0}^{\mathrm{ES}} + D_{0}^{\mathrm{UE}} = 0.4641$ [ms] \\ 
            Bandwidth & $1$ [MHz]                                 & Energy consumption model &$f^{\mathrm{ES}} = 1/0.0774$ [hJ/TFLOPs]\\
            Size of latent feature & $(d^{\mathrm{C}}, d^{\mathrm{W}}, d^{\mathrm{H}}) = (16, 64, 64)$ & $\quad$ &$f^{\mathrm{UE}} =0.8f^{\mathrm{ES}}$ [hJ/TFLOPs]\\
            Size of action particles & $K = 32$ & \quad & $E_{0}^{\mathrm{ES}} + E_{0}^{\mathrm{UE}} = 0.7320$ [hJ] \\
            Learning rate & $\nu_{\zeta} = \nu_{\lambda} = 0.02$& FLOPs of single U-Net block & $O^{\mathrm{E}} = 0.2200$ [TFLOPs]\\
            \quad&  $\nu_{\theta} = 0.0005$ & FLOPs of VAE encoder& $O^{\mathrm{UNet}} = 11.2482$ [TFLOPs]\\
            M-step/E-step iterations & $I_{\mathrm{E}}=10$, $I_{\mathrm{M}}=6$  &  FLOPs of VAE decoder & $O^{\mathrm{D}} = 10.2310$ [TFLOPs]\\
            \bottomrule
        \end{tabular}\label{table_param}
    }
\end{table}

\begin{figure}[!htbp]
    \vspace{-2.5em}
    \centering
    \includegraphics[width=0.65\textwidth]{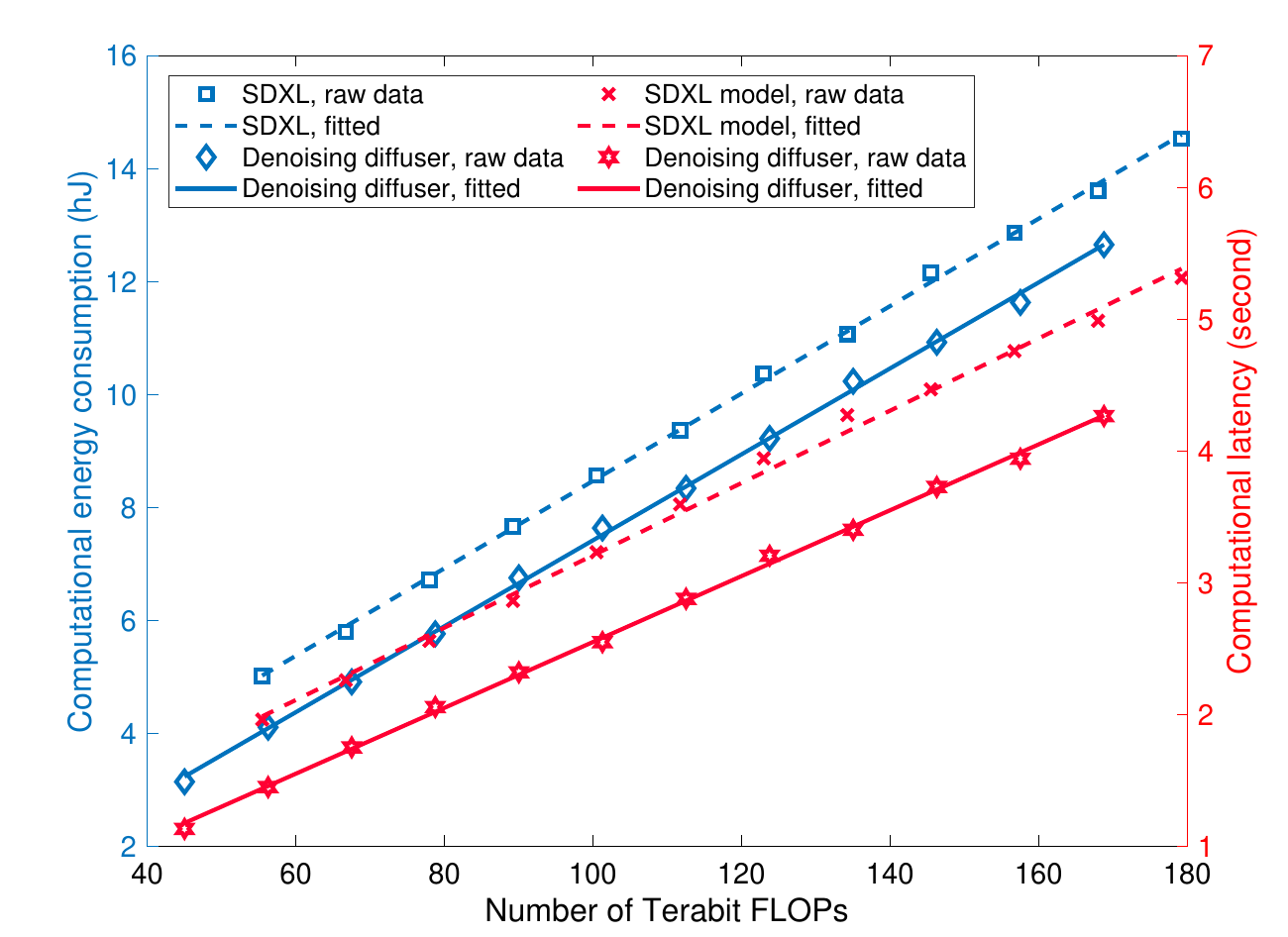}
    \caption{Latency and energy consumption evaluation of the proposed MEG framework.}
    \label{fig_fit_model}
    \vspace{0.5em}
\end{figure}

For feature transmissions of MEG, we set the transmitting power at BS at $P_{0}=30$ dBm, and the channel bandwidth allocated to the UE is $1$ MHz. 
For decoding images from the received features over the noisy channels, the peak signal-to-noise ratio (PSNR) is given by $10\log_{10}\frac{P_{0}}{\sigma^2}=10$ dB. 
We collect a subset of high-resolution images from  LAIONON-COCO-Aesthetic dataset \cite{LAIONON_COCO} with $20000$ training samples and $2000$ test samples. 
The image quality is evaluated by calculating the mean square error (MSE) between the generated images and the reference images, 
where the reference image is generated by the centralized \textit{SDXL} with $L_{\max}=12$ and transmitted through perfect channels. 
Each floating point number takes up $16$ bits during both computations and transmissions. 
Based on the measurement results in a Linux workstation with $2$ Nvidia A40 GPUs and PyTorch,  
both realistic latency and energy consumption are linearly related to FLOPs, as shown in Fig. \ref{fig_fit_model}. 
This validates the soundness of the considered latency and energy consumption models. 
The corresponding simulation parameters can be summarized in Table \ref{table_param}.

To verify the performance of the proposed framework, we introduce the following baselines. 
\begin{itemize}
    \item \textbf{Centralized Generation} (perfect channels): The centralized \textit{SDXL} model is depolyed at the ES, 
    which takes $L_{\max} = 12$ denoising steps. The generated $1024\times 1024$ colour images will be transmitted to the UE through the perfect channels. 
    \item \textit{\textbf{MEG-Split}}: The pretrained \textit{SDXL} model is directly split into two parts and deployed at the ES and UE without finetuning and feature compression. 
    \item \textbf{MEG-Split-Finetune}: The split \textit{SDXL} sub-models are deployed at the ES and UE, and the decoder is fine-tuned over noisy channels to minimize MSE for distortion mitigation. 
    \item \textbf{MEG-Pruning}: Based on the split \textit{SDXL} sub-models, a pruning-based adaptive feature encoder scheme similar to \cite{FeatureEncoding} is utilized, which can dynamically compress transmitted features in the online channel environment.
    \item \textbf{MEG-PPO-Lagrangian}: 
    The Lagrangian-based constrained learning method is utilized for online prediction of MEG, which is integrated into the proximal policy optimization (PPO) framework \cite{PPO}. 
    To enhance the dual variable update stability of the vanilla Lagrangian-based method, the proportional and derivative (PID) control is also considered \cite{PID}. 
\end{itemize}
To ensure fairness, a pretrianed few-step denoising diffuser is enabled at all baselines. 
To further accelerate computations for \textit{MEG-Pruning} and our proposed MEG scheme, we also implement the token merging for the computation 
of each self-attention layer with a token merging ratio $\widetilde{\beta}=0.5$.

\begin{table*}[!b]
    \vspace{2em}
    \centering
    \caption{Detailed numerical result comparison of different methods. $L_{\max}=12$.}
    \resizebox{1\textwidth}{!}{ 
        \begin{tabular}{c|cccccc}
            \toprule
            {Method\/} & Centralized Generation & {\textit{MEG-Split}} & {\textit{MEG-Split-Finetune}} & \textit{MEG-Pruning}& MEG (proposed) & MEG (proposed)\\
            Performance&(Perfect channels, $L=12$)&($L=8$)&($L=8$)&($L=8,\beta=0.1)$ &($L=8, \beta=0.5$)&($L=4, \beta=0.5$)\\
            \hline
            MSE & 0 (Reference) & 0.2842 & 0.0106 & 0.0153 & 0.0095 & 0.0246\\ \hline
            Transmission & \multirow{2}*{60.4680} & \multirow{2}*{1.2598} & \multirow{2}*{1.2598} & \multirow{2}*{1.1337} &  \multirow{2}*{0.7547}&  \multirow{2}*{0.7547}\\
            latency (second) & & & & & \\\hline
            Computation & \multirow{2}*{7.5806} & \multirow{2}*{5.2122} & \multirow{2}*{5.2122} & \multirow{2}*{3.2316} &  \multirow{2}*{3.2316} &\multirow{2}*{2.0287}\\
            latency (second) & & & & & \\
            \bottomrule
        \end{tabular}\label{table_result}
    }
\end{table*}

\subsection{Numerical Results}

{
\begin{figure*}[!tbp]
    \vspace{-2.3em}
    \centering
    \includegraphics[width=0.98\textwidth]{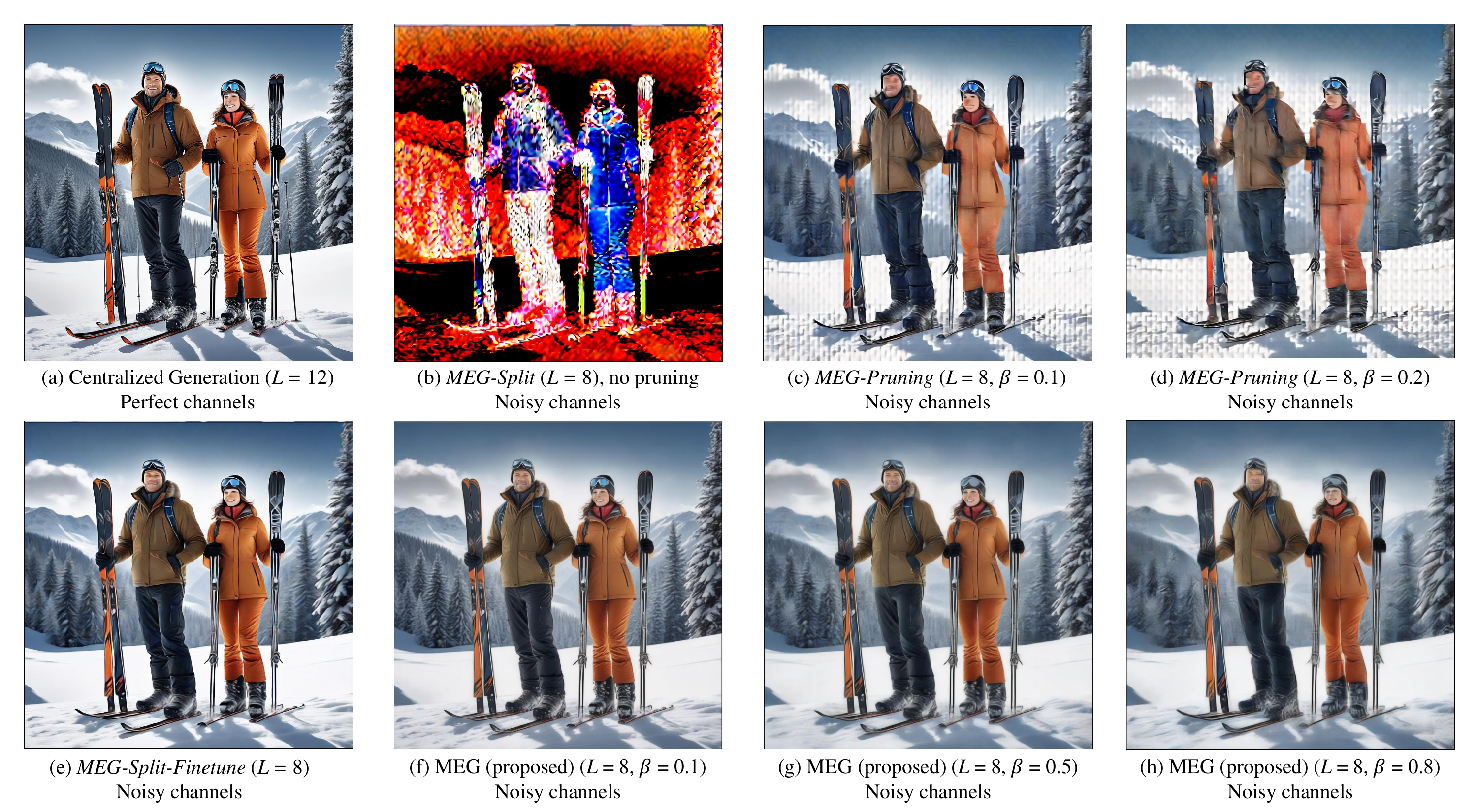}
    \caption{Images created by different generation methods. The text prompt is given by ``Two people standing in the snow with skis''.}
    \label{fig_img_sample}
    \vspace{0.5em}
\end{figure*}
}

\subsubsection{Offline Distillation Performance}
Exploiting the proposed dynamic diffusion and feature merging scheme, we train the backbone architecture of distributed sub-models via offline distillation. 
Given the same prompt, Fig. \ref{fig_img_sample} compares the image samples generated by different models. 
The detailed numerical results have been presented in Table \ref{table_result}.  
As shown in Fig. \ref{fig_img_sample}, the \textit{MEG-Split} scheme suffers from highly noisy images. 
As an advance, \textit{MEG-Split-Finetune} can overcome the channel noises and generates high-fidelity images, but still requires a relatively high latency.  
In contrast, \textit{MEG-Pruning} can further reduce the computation and transmission latencies, but leads to inevitable distortionas some image details are missing.
Based on feature merging, our proposed MEG scheme can reduce over $90\%$ and over $40\%$ latency compared to the centralized generation and \textit{MEG-Split-Finetune} schemes, respectively, 
while generating meaningful high-resolution images and retain sufficient image details. 

\begin{figure}[!tbp]
    \vspace{-2.3em}
    \centering
    \includegraphics[width=0.65\textwidth]{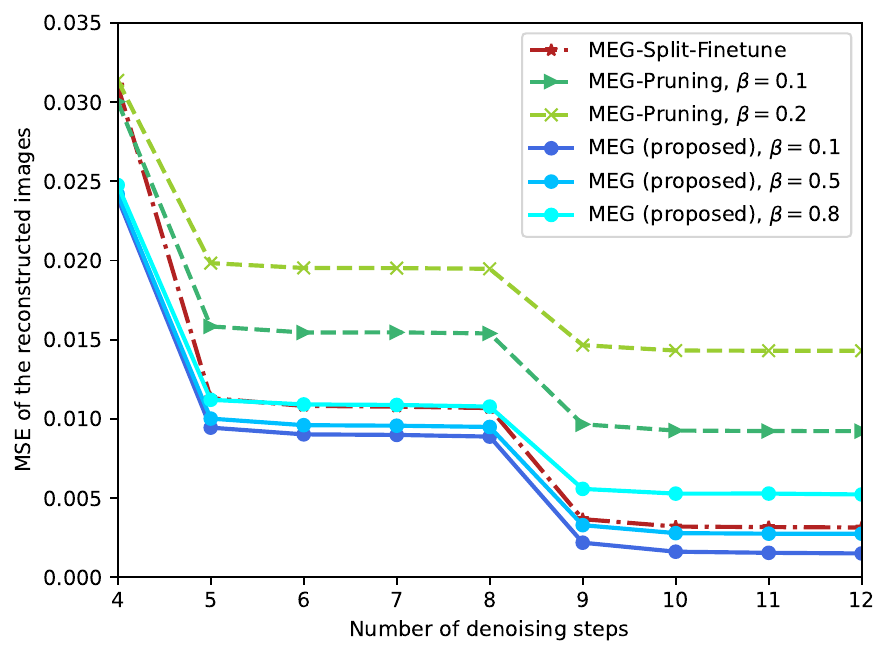}
    \caption{Comparisons of average MSE achieved by different generation schemes.}
    \label{fig_MSE_offline}
    \vspace{1em}
\end{figure}

\begin{figure}[!htbp]
    \vspace{-2.5em}
    \centering
    \subfloat[]{\includegraphics[width=4in]{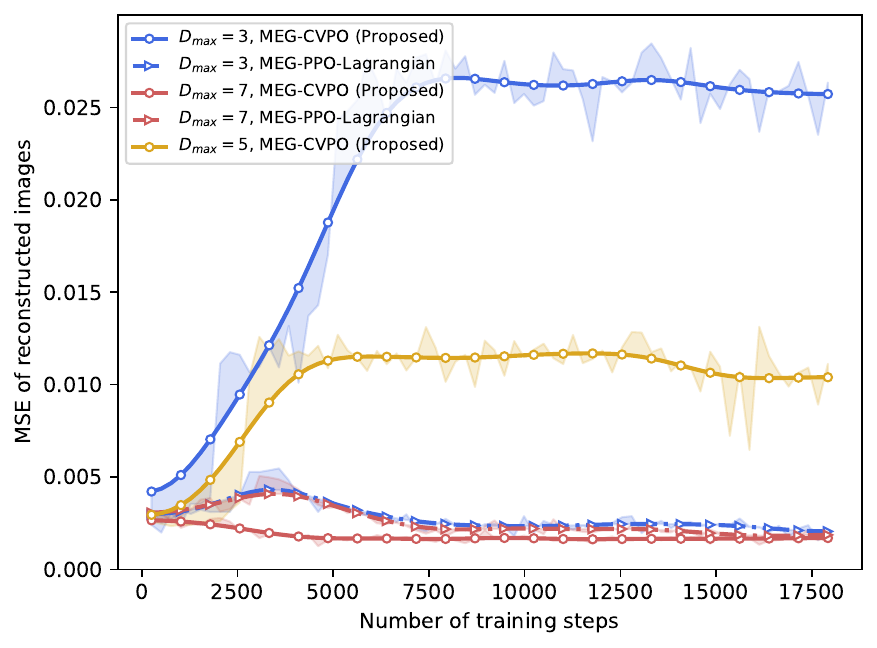}%
    \label{fig_MSE_CVPO}}
    \\
    \subfloat[]{\includegraphics[width=4in]{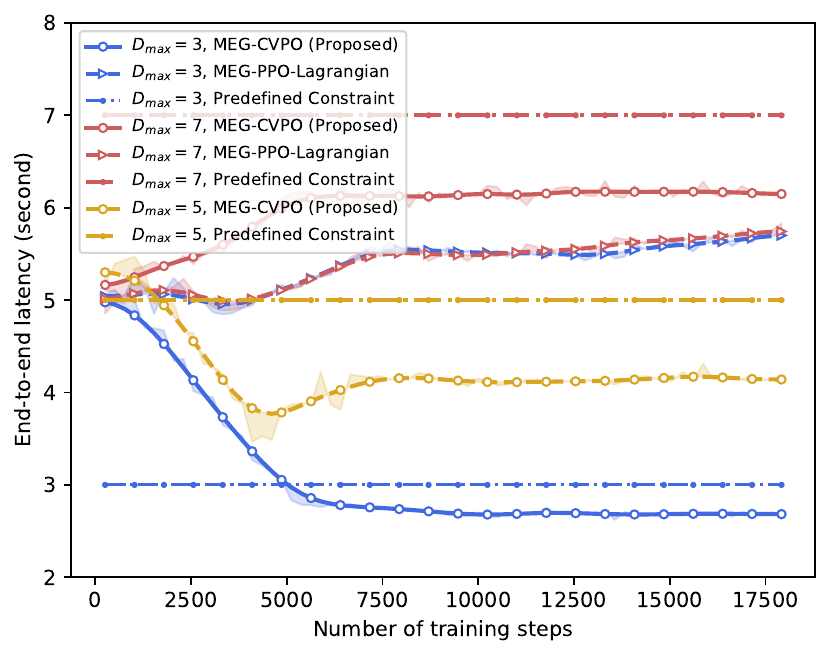}%
    \label{fig_latency_CVPO}}
    \caption{Performance comparisons of constrained learning algorithms in the test stage. $E_{\max} = 1.8$ kJ. 
    (a) Average MSE of the generated images. 
    (b) Performance of latency constraint guarantees.}\vspace{0.6em}
    \label{fig_MSE_latency_CVPO}
\end{figure}

Fig. \ref{fig_MSE_offline} demonstrates the performance of different generation schemes under various options of different denoising steps and feature compressions. 
Generally, the MSE decreases with denoising steps and increases with feature compression ratios, which confirms the trade-off between the image generation quality and costs. 
It can be observed that the \textit{MEG-Pruning} scheme results in a much higher MSE than \textit{MEG-Split-Finetune} scheme, even when the pruning ratio is low. 
In contrast, the proposed dynamic feature merging scheme generates high-quality images 
comparable to \textit{MEG-Split-Finetune} the when feature merging ratio $\beta = 0.5$. 
This verifies the feature compression efficiency of the proposed scheme for distributed image generations.  
It is worth noting that the proposed scheme may 
outperform \textit{MEG-Split-Finetune} when the denoising step and the feature merging ratio is small. 
This implies that training with length-variable merged features may enhance the noise resistance performance of the feature decoder at the UE.

\begin{figure}[!htbp]
    \vspace{-2.5em}
    \centering
    \includegraphics[width=4in]{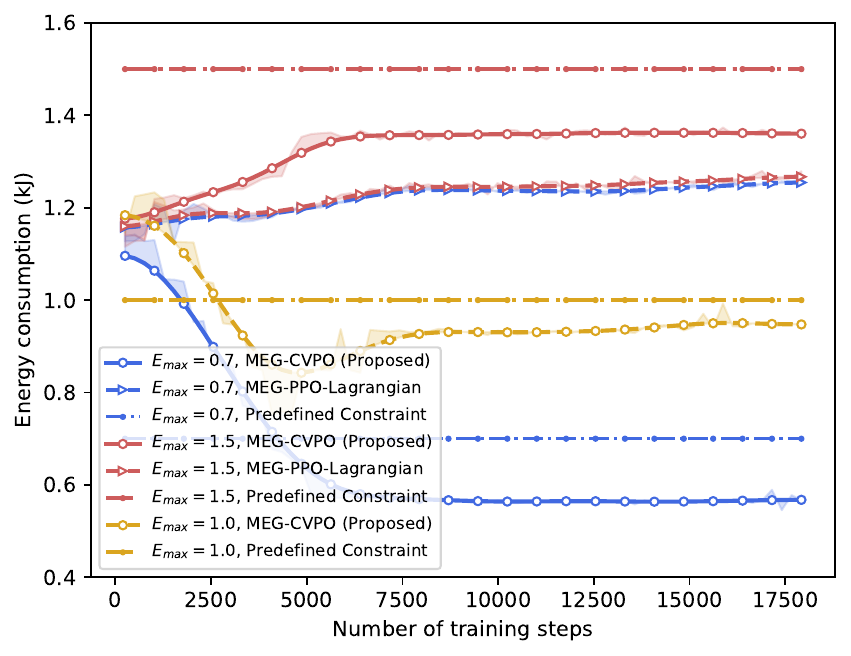}
    \caption{Performance comparisons of energy constraint guarantees. } \vspace{0.6em}
    \label{fig_energy_constraint}
\end{figure}

\subsubsection{Constrained Learning Performance}
We further validate the constrained learning performance of the proposed MEG-CVPO algorithm 
for online prediction. 
The batch size for policy learning is set as $64$. 
We train both MEG-CVPO and \textit{MEG-PPO-Lagrangian} algorithms in $100$ epochs, and utilize the Adam optimizer for gradient descent. 
Fig. \ref{fig_MSE_latency_CVPO} presents the average MSE and latency guarantee performance for different constrained learning algorithms. 
Both MEG-CVPO and \textit{MEG-PPO-Lagrangian} can converge within $15000$ training steps. 
As shown in Fig. \ref{fig_MSE_latency_CVPO}(a), when the maximum latency requirement is relaxed (i.e., $D_{\max} = 7$ seconds), 
MEG-CVPO can achieve a higher generation quality with a faster convergence speed compared to  \textit{MEG-PPO-Lagrangian}, 
which verifies its efficiency in optimizing denoising steps and feature merging ratios. 
Both MEG-CVPO and \textit{MEG-PPO-Lagrangian} can guarantee the latency constraints in this case, as demonstrated in Fig. \ref{fig_MSE_latency_CVPO}(b). 
However, 
when a low latency is required  (i.e., $D_{\max} = 3$ seconds), MEG-CVPO leads to a much higher MSE than \textit{MEG-PPO-Lagrangian}, 
since the budgets of computational and transmission overheads are significantly reduced. 
As a result, the proposed MEG-CVPO algorithm can still stringently guarantee the latency constraint, 
while \textit{MEG-PPO-Lagrangian} suffering obvious constraint violations. 
Furthermore, the proposed MEG-CVPO algorithm can also adaptively adjust the dynamic denoising steps and feature merging ratios to 
strictly satisfy a latency requirement of $D_{\max} = 5$ seconds, which will yield a compromised image quality. 
This also demonstrates that the proposed MEG-CVPO algorithm can realize a controllable quality-latency trade-off for on-device image generation.

Fig. \ref{fig_energy_constraint} further compares the energy consumptions of different algorithms under different  energy budgets. 
We set $D_{\max} = 8$ seconds and $E_{\max}=$here.
The \textit{MEG-PPO-Lagrangian} algorithm severely violates the energy consumption constraint when the energy budget is small. 
In contrast, by optimizing the feasible policy distributions and learning policies in the corresponding trusted regions, 
the proposed MEG-CVPO algorithm can strictly satisfy various energy consumption requirements. 
In addition, the gap between the energy consumption of MEG-CVPO and the specified energy budget is tighter compared to that of \textit{MEG-PPO-Lagrangian}. 
This also indicates that MEG-CVPO can improve the efficiency of resource utilization in resource-constrained systems.

\section{Conclusion}
A novel accelerated mobile edge generation (MEG) framework has been proposed to enable cost-efficient AIGC services at edge networks. 
Leveraging the decomposed LDM sub-models distributed across the ES and the UE, the proposed framework achieves on-device high-resolution image generation with low transmission and computing overheads.
The denoising steps and feature merging ratio were jointly optimized, aiming to maximize the image generation quality while satisfying latency and energy consumption constraints.
To address this optimization problem and tailor the LDM sub-models for dynamic acceleration, 
we developed a low-complexity dynamic acceleration MEG protocol. 
More specifically, a backbone architecture has been trained via offline distillation. 
Then, the dynamic diffusion and feature merging were realized in the online channel environment, which can be modelled as a constrained MDP. 
An MEG-CVPO algorithm has been further developed, which effectively trained the policy to enhance image quality and ensure feasible policy distributions.
Numerical results revealed the efficiency of the proposed framework in accelerating high-resolution image generation and resisting distortions. 
Additionally, the proposed MEG-CVPO algorithm can effectively ensure constraints and adaptively balance between generation quality and costs. 
For the future outlook, practical and scalable edge resource management can be further explored by extending the proposed framework into parallel and multi-user AIGC applications.


\begin{thebibliography}{1}

\bibitem{ChatGPT}
O. AI, ``Chatgpt: Optimizing Language Models for Dialogue,'' 
``Accessed Jun. 30, 2024'', [Online]. Available: https://openai.com/blog/chatgpt/.

\bibitem{DALLE}
O. AI, ``DALL-E 3,'' 
https://openai.com/index/dall-e-3/, [Online]. Available: https://openai.com/index/dall-e-3/. 


\bibitem{GAI} S. Huang, P. Grady, and GPT-3, ``Generative AI: A Creative New World,''
``Accessed Jun. 30, 2024'', [Online]. Available: https://www.sequoiacap.
com/article/generative-{AI}-a-creative-new-world/.

\bibitem{Mobile_GAI} M. Xu et al., ``Unleashing the Power of Edge-Cloud Generative AI in Mobile Networks: A Survey of AIGC Services,'' 
\textit{IEEE Commun. Surveys Tut.}, vol. 26, no. 2, pp. 1127-1170, Secondquarter 2024. 

\bibitem{AIGC_SematicCom} 
R. Cheng, Y. Sun, D. Niyato, L. Zhang, L. Zhang, and M. A. Imran, 
``A Wireless AI-Generated Content (AIGC) Provisioning Framework Empowered by Semantic Communication,''
\textit{arXiv preprint arXiv: 2310.17705}, 2024.

\bibitem{DeviceOnlyInference}
H. Cai, C. Gan, T. Wang, Z. Zhang, and S. Han, ``Once-for-all: Train
One Network and Specialize it for Efficient Deployment,'' in 
\textit{Proc. Int. Conf. Learn. Represent.}, Addis Ababa, Ethiopia, Apr. 2020.

\bibitem{MEG}
R. Zhong, X. Mu, Y. Zhang, M. Jabor, and Y. Liu, 
``Mobile Edge Generation: A New Era to 6G,''
\textit{IEEE Netw.}, early access, 2024. 

\bibitem{Edge_Coinference_Li}
E. Li, Z. Zhou, and X. Chen, 
``Edge Intelligence: On-Demand Deep Learning Model Co-Inference with Device-Edge Synergy,'' 
in \textit{Proc. Workshop Mobile Edge Commun.}, Budapest, Hungary, Aug. 2018, pp. 31-36.

\bibitem{EdgeAI_Shi}
Y. Shi, K. Yang, T. Jiang, J. Zhang, and K. B. Letaief, 
``Communication-Efficient Edge AI: Algorithms and Systems,'' 
\textit{IEEE Commun. Surveys Tuts.}, vol. 22, no. 4, pp. 2167-2191, 4th Quart., 2020.

\bibitem{EdgeAI_Xing}
H. Xing, G. Zhu, D. Liu, H. Wen, K. Huang and K. Wu, 
``Task-Oriented Integrated Sensing, Computation and Communication for Wireless Edge AI,'' \textit{IEEE Netw.}, vol. 37, no. 4, pp. 135-144, July/August 2023.


\bibitem{Edge_Coinference_Kang}
Y. Kang, J. Hauswald, C. Gao, A. Rovinski, T. Mudge, J. Mars, and
L. Tang, ``Neurosurgeon: Collaborative Intelligence between the Cloud and Mobile Edge,'' 
\textit{ACM SIGARCH Comput. Archit. News}, 
vol. 45, no. 1, pp. 615-629, Apr. 2017.

\bibitem{Edge_Coinference_Spliting}
H. Li, C. Hu, J. Jiang, Z. Wang, Y. Wen, and W. Zhu, ``JALAD: Joint
Accuracy- and Latency-Aware Deep Structure Decoupling for Edge-Cloud Execution,'' 
in \textit{Proc. Int. Conf. Parallel Distrib. Syst.}, Singapore, Dec.
2018, pp. 671-678.



\bibitem{FeatureEncoding}
J. Shao, Y. Mao and J. Zhang, 
``Learning Task-Oriented Communication for Edge Inference: An Information Bottleneck Approach,'' 
\textit{IEEE J. Sel. Areas Commun.}, vol. 40, no. 1, pp. 197-211, Jan. 2022. 

\bibitem{InformationBottleneck}
N. Tishby, F. C. Pereira and W. Bialek, ``The Information Bottleneck Method,'' 
\textit{Proc. Annu. Allerton Conf. Commun. Control Comput.}, pp. 368-377, Oct. 1999. 

\bibitem{Bottlenet}
J. Shao and J. Zhang, ``Bottlenet++: An End-to-End Approach for Feature Compression in Device-Edge Co-Inference Systems,'' 
in \textit{Proc. Int. Conf. Commun. Workshop}, Dublin, Ireland, Jun. 2020, pp. 1-6.


\bibitem{MEG_DT}
X. Xu, R. Zhong, X. Mu, Y. Liu, K. Huang, 
``Mobile Edge Generation-Enabled Digital Twin: Architecture Design and Research Opportunities,''
\textit{arXiv preprint arXiv: 2407.02804}, 2024.

\bibitem{DistributedDiff}
H. Du, R. Zhang, D. Niyato, J. Kang, Z. Xiong, D. I. Kim, X. Shen, and H. V. Poor, 
``User-Centric Interactive AI for Distributed Diffusion Model-based AI-Generated Content,''
\textit{arXiv preprint arXiv: 2311.11094}, 2023. 

\bibitem{AIGC_Pricing}
J. Wang, H. Du, D. Niyato, J. Kang, Z. Xiong, D. Rajan, S. Mao et al., ``A Unified Framework for Guiding Generative AI with Wireless
Perception in Resource Constrained Mobile Edge Networks,'' 
\textit{arXiv preprint arXiv:2309.01426}, 2023.

\bibitem{TrustWorthAIGC}
S. Li, Xi Lin, Y. Liu, and J. Li, 
``Trustworthy AI-Generative Content in Intelligent 6G Network: Adversarial, Privacy, and Fairness'', 
\textit{arXiv preprint arXiv: 2405.05930}, 2024. 

\bibitem{UNet}
O. Ronneberger, P. Fischer, and T.Brox, 
``U-Net: Convolutional Networks for Biomedical Image Segmentation,'' 
\textit{arXiv preprint arXiv:1505.04597}, 2015.

\bibitem{denoising}
C. Meng, Y. Song, J. Song, J. Wu, J. Zhu, and S. Ermon. 
``SDEdit: Guided Image Synthesis and Editing with Stochastic Differential Equations,'' 
\textit{arXiv preprint, arXiv:2108.01073}, 2021.

\bibitem{DDIM}
J. Song, C. Meng, and S. Ermon, ``Denoising Diffusion Implicit Models,'' 
\textit{arXiv preprint, arXiv:2010.02502},  June 2022.

\bibitem{Energy_FLOPs}
R. Desislavov, et al., ``Compute and Energy Consumption Trends in Deep Learning Inference,'' 
\textit{arXiv preprint arXiv:2109.05472}, 2021.

\bibitem{Token_Merging}
D. Bolya, C. Fu, X. Dai, P. Zhang, C. Feichtenhofer, and J. Hoffman, ``Token Merging: Your ViT But Faster,'' 
in \textit{Proc. Int. Conf. Learn. Represent.}, 2023. 

\bibitem{Token_Merging_SD}
D. Bolya, J. Hoffman, ``Token Merging for Fast Stable Diffusion,'' 
in \textit{Proc. IEEE/CVF Conf. Comput. Vis. Pattern Recognit. (CVPR)}, 2023, pp. 4599-4603.

\bibitem{PeRFlow}
H. Yan, X. Liu, J. Pan, J. H. Liew, Q. Liu, and  J. Feng, 
``Perflow: Piecewise Rectified Flow as Universal Plug-and-Play Accelerator'', 
\textit{arXiv preprint arXiv:2405.07510}, May 2024.

\bibitem{Lagrangian}
D. Ding, k. Zhang, t. Basar, and M. Jovanovic, 
``Natural Policy Gradient Primal-Dual Method for Constrained Markov Decision Processes,''
in \textit{Proc. Adv. Neural Inf. Process. Syst.}, pp. 8378-8390, 2020.

\bibitem{CVPO}
Z. Liu, Z. Cen, V. Isenbaev, W. Liu, S. Wu, B. Li, and D. Zhao,  
``Constrained Variational Policy Optimization for Safe Reinforcement Learning,''
in \textit{Proc. Int. Conf. Mach. Learn.}, pp. 13644-13668, 2022.

\bibitem{RL_optimality}
S. Levine, ``Reinforcement Learning and Control as Probabilistic Inference: Tutorial and Review,'' 
\textit{arXiv preprint arXiv:1805.00909}, 2018.

\bibitem{SDXL}
D. Podell, Z. English, K. Lacey, A. Blattmann, T. Dockhorn, J. Muller, J. Penna, and R. Rombach, 
``SDXL: Improving Latent Diffusion Models for High-Resolution Image Synthesis,''
\textit{arXiv preprint arXiv:2307.01952}, 2023.


\bibitem{LAIONON_COCO}
C. Schuhmann, R. Beaumont, R. Vencu, C. Gordon, R. Wightman, M. Cherti, T. Coombes, A. Katta,
C. Mullis, M. Wortsman, et al. ``Laion-5b: An Open Large-Scale Dataset for Training Next Generation
Image-Text Models,'' 
\textit{Proc. Adv. Neural Inf. Process. Syst.}, pp. 25278-25294, 2022.

\bibitem{PPO}
J. Schulman, F. Wolski, P. Dhariwal, A. Radford, and O. Klimov. 
``Proximal Policy Optimization Algorithms,'' 
\textit{arXiv preprint arXiv:1707.06347}, 2017.


\bibitem{PID}
A. Stooke, J. Achiam, and P. Abbeel, 
``Responsive Safety in Reinforcement Learning by PID Lagrangian Methods,'' 
in \textit{Int. Conf. Mach. Learn.}, pp. 9133-9143, 2020.

\end{thebibliography}
\end{document}